\documentclass[10pt, letterpaper]{IEEEtran}
\usepackage{amsmath,amsthm}
\usepackage{tabularx,multirow}
\usepackage{graphicx,subfigure,comment}
\usepackage[table]{xcolor}
\usepackage{url} %for \path
\usepackage{dsfont}
\usepackage{cite}
\usepackage{siunitx}
\usepackage{physics} %dv
\usepackage{pstool}
\usepackage{paralist}
\usepackage{enumitem}
\usepackage{hyperref}
\usepackage{autonum} %hyperref before autonum or else arxiv makes a mess
\usepackage{algpseudocode}
\usepackage[Algorithm]{algorithm}

\setlist[itemize]{leftmargin=*}
\setlist[enumerate]{leftmargin=*}

\newcommand{\Fig}[1]{Fig.~\ref{fig:#1}}
\newcommand{\Prop}[1]{Property~\ref{prop:#1}}
\newcommand{\Ex}[1]{Example~\ref{ex:#1}}
\newcommand{\Lem}[1]{Lemma~\ref{lem:#1}}
\newcommand{\Thm}[1]{Theorem~\ref{thm:#1}}

\newcommand{\Sec}[1]{Sec.~\ref{sec:#1}}
\newcommand{\Tab}[1]{Tab.~\ref{tab:#1}}
\newcommand{\Eq}[1]{(\ref{eq:#1})}

\newcommand{\ind}[1]{\mathds{1}_{#1}}

\newcommand{\Vc}{\mathcal{V}}

\newcommand{\Mc}{\mathcal{M}}
\newcommand{\Sc}{\mathcal{S}}
\newcommand{\PP}{\mathds{P}} % probability
\newtheorem{theorem}{Theorem}
\newtheorem{property}{Property}

\newtheorem{lemma}{Lemma}
\newtheorem{example}{Example}
\newtheorem*{example*}{Example}

\allowdisplaybreaks

\begin{document}

\title{Reducing Service Deployment Cost\\Through VNF Sharing 
} %title

\author{
Francesco~Malandrino,~\IEEEmembership{Senior~Member,~IEEE,}
Carla Fabiana~Chiasserini,~\IEEEmembership{Fellow,~IEEE,}
Gil~Einziger,~\IEEEmembership{Member,~IEEE,}
Gabriel~Scalosub,~\IEEEmembership{Senior~Member,~IEEE}% <-this % stops a space
\IEEEcompsocitemizethanks{\IEEEcompsocthanksitem F.~Malandrino and C.-F.~Chiasserini are with CNR-IEIIT, Italy. C.-F. Chiasserini is with Politecnico di Torino, Italy. G.~Einziger and G.~Scalosub are with Ben-Gurion University of the Negev, Israel.
\IEEEcompsocthanksitem This work was supported by the EU Commission through
the 5GROWTH project (grant agreement no. 856709). The work of G.~Scalosub has been supported by the Israel Science Foundation
(grant No. 1036/14) and the Neptune Consortium, administered by the Israeli Ministry
of Economy and Industry.
%\IEEEcompsocthanksitem A preliminary version~\cite{noi-wowmom19} of this work has appeared at the IEEE WoWMoM 2019 conference.
}% <-this % stops an unwanted space
}

\maketitle

\begin{abstract}
%Upcoming 5G networks will have both computational and forwarding capabilities, which enable them to support multiple third-party (``vertical'') services with the same infrastructure. Such services are described as graphs of virtual (network) function (VNFs), and it is often the case that one VNF is common across multiple services. In this context, we study the {\em VNF sharing} problem, requiring to decide (i) whether a certain VNF shall be shared among multiple services, i.e., if doing so is beneficial, (ii) how to scale the virtual machines running the shared VNFs, in order to cope with the combined load of the corresponding services, and (iii) how to assign the priorities to services within shared VNFs. All decisions are made with the aim of minimizing the cost for the mobile operator, subject to requirements on end-to-end service performance, e.g., total delay. Importantly, we show that service priorities should not be determined {\em a priori}, e.g., by verticals; rather, they should be managed dynamically and allowed to change across VNFs. We propose an effective and efficient methodology, called FlexShare, able to reap the benefits of such additional flexibility by making near-optimal VNF-sharing decisions in polynomial time and within a constant factor from the optimum. After analyzing the complexity of FlexShare and its competitive ratio, we evaluate its performance using real-world VNF graphs, finding that FlexShare consistently outperforms baseline solutions using per-service priorities.

Thanks to its computational and forwarding capabilities, the mobile network infrastructure can support  several third-party (``vertical'') services, each composed of a graph of virtual (network) functions (VNFs).  Importantly, one or more VNFs are often common to multiple services, thus the services deployment cost could be reduced by letting the services share the same VNF instance instead of devoting a separate instance to each service. By doing that, however, it is critical that the target KPI (key performance indicators) of all services are met. To this end, we  study the {\em VNF sharing} problem and make decisions on 
\begin{inparaenum}[(i)]
\item when sharing  VNFs among multiple services is possible,
\item how to adapt the virtual machines running the shared VNFs to the combined load of the assigned services, and
\item how to prioritize the services traffic within shared VNFs. 
\end{inparaenum}
All decisions aim to minimize the cost for the mobile operator, subject to requirements on end-to-end service performance, e.g., total delay. Notably, we show that the aforementioned priorities should be managed dynamically and vary across VNFs. We then propose the FlexShare algorithm to provide near-optimal VNF-sharing and priority assignment decisions in polynomial time. We prove that FlexShare is within a constant factor from the optimum and, using real-world VNF graphs, we show that it consistently outperforms baseline solutions.
\end{abstract}

\section{Introduction}
\label{sec:intro}

Emerging mobile networks do not  only forward data, but they can also process the data:  based on the software-defined networking (SDN) and the network function virtualization (NFV) paradigms, they run virtual (network) functions (VNFs) and perform network-related (e.g., firewalls) or service-specific (e.g., video transcoding) tasks. Such processing at the edge of the network infrastructure  allows for lower latency, higher efficiency, and lower costs compared to current cloud-based architecture. 
These new capabilities of network systems bring forth a new relationship between mobile network operators (MNOs) and third-party companies (``verticals'') providing the services. On the one hand, verticals make business agreements with MNOs, specifying
\begin{inparaenum}[(i)]
\item the service they wish to run, defined by a VNF graph, i.e., the set of VNFs composing the service, properly connected to each other; 
\item the end-to-end target key performance indicators (KPIs), e.g., throughput, delay, or reliability for each service. 
\end{inparaenum}
On the other hand,  MNOs seek to maintain the target  KPIs of the deployed services while minimizing their deployment cost.
Inefficiently-used infrastructure can indeed result into significant costs for the MNO, thus jeopardizing its revenue: as an example, \cite{meisner2009powernap} reports that idle servers consume 60\%~of their peak power.

To efficiently support multiple services, the  {\em network slicing} paradigm for backhaul infrastructure has been  introduced~\cite{slicing}. Under this paradigm, the mobile operator's backhaul infrastructure, e.g., routers and servers, supports services from different verticals, while guaranteeing  isolation  and honoring the target KPIs of each service. Network slicing also supports composed services (i.e., services whose VNF graphs include sub-graphs, each corresponding to a child service~\cite{slicing2}), thus enabling the corresponding slices to include common sub-slices~\cite{5gppp-architecture,ietf-mano}. A typical example of a child service is the cellular Evolved Packet Core (EPC)~\cite{slicing2}, which is a common component of the mobile services required by the verticals.

Upon creating a slice, MNOs must
\begin{inparaenum}[(i)]
\item assign it the necessary resources (e.g., virtual machines and virtual links connecting them), and
\item decide which VNFs to run at each host.
\end{inparaenum}
The latter problem, known as the VNF placement problem, is often formulated as a cost minimization problem subject to the target KPIs. Importantly, significant cost savings can be achieved by {\em sharing} individual VNFs or sub-slices among services, whenever possible.
The vast majority of VNF placement studies~\cite{placement-cinesi,placement-infocom,noi-infocom18,PlacementInfocom2019} consider scenarios where all placement decisions are made by a centralized entity, often the NFV Orchestrator (NFVO) in the ETSI Management and Orchestration (MANO) framework\cite{etsimano,ietf-mano}.
Also, such an entity is in the position to make fine-grained decisions on the usage of individual hosts and links.

\begin{figure}
\includegraphics[width=1\columnwidth]{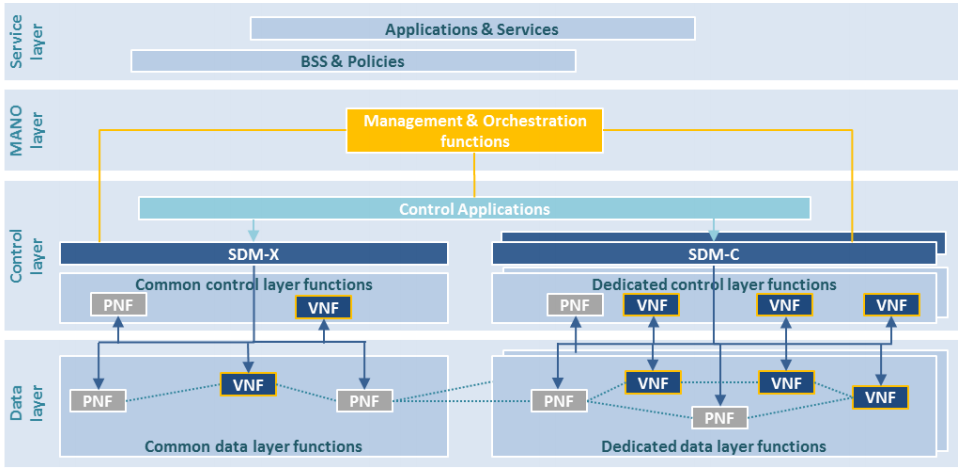}
\caption{
   Architectural view of 5G networks according to 5G-PPP. Source:~\cite{slicing2}.
    \label{fig:5gppp}
} %caption
\end{figure}

However, such a scenario is not typical of real-world implementations. Indeed, ETSI~\cite[Sec.~8.3.6]{ifa7} specifies four granularity levels for placement decisions:
\begin{itemize}
    \item individual host;
    \item zone (i.e., a set of hosts with certain common features);
    \item zone group;
    \item point-of-presence (PoP), e.g., a datacenter.
\end{itemize}
Real-world mobile networks implementations, including~\cite{pimrc-wp4,norma,5gt}, assume that the NFVO, or similar entities, make PoP-level decisions. Placement and sharing decisions within individual PoPs, instead,  can be made by other entities under different names and with slight variations between IETF~\cite[Sec.~3]{ietf-mano}, the NGMN alliance~\cite[Sec.~8.9]{ngmn-mano}, and 5G-PPP~\cite[Sec.~2.2.2]{5gppp-architecture}. Without loss of generality, in this work we focus on the last architecture which includes a Software-Defined Mobile Network Coordinator (SDM-X) illustrated in \Fig{5gppp}.  The SDM-X operates at a lower level of abstraction than the NFVO and makes intra-PoP VNF placement and sharing decisions. 
Specifically, for each newly-requested service, the SDM-X has to make decisions on:
\begin{itemize}
    \item whether any of the VNFs requested by the new service shall be provided through existing instances thereof;
    \item if so, how to assign the priorities to the traffic flows of the different services using the same VNF instance;
    \item if not, which virtual machine (VM) to instantiate as a new VNF instance;
    \item how to adjust (e.g., scale up/down) the computational capability of VMs within the PoP.
\end{itemize}

\begin{figure}
\includegraphics[width=1\columnwidth]{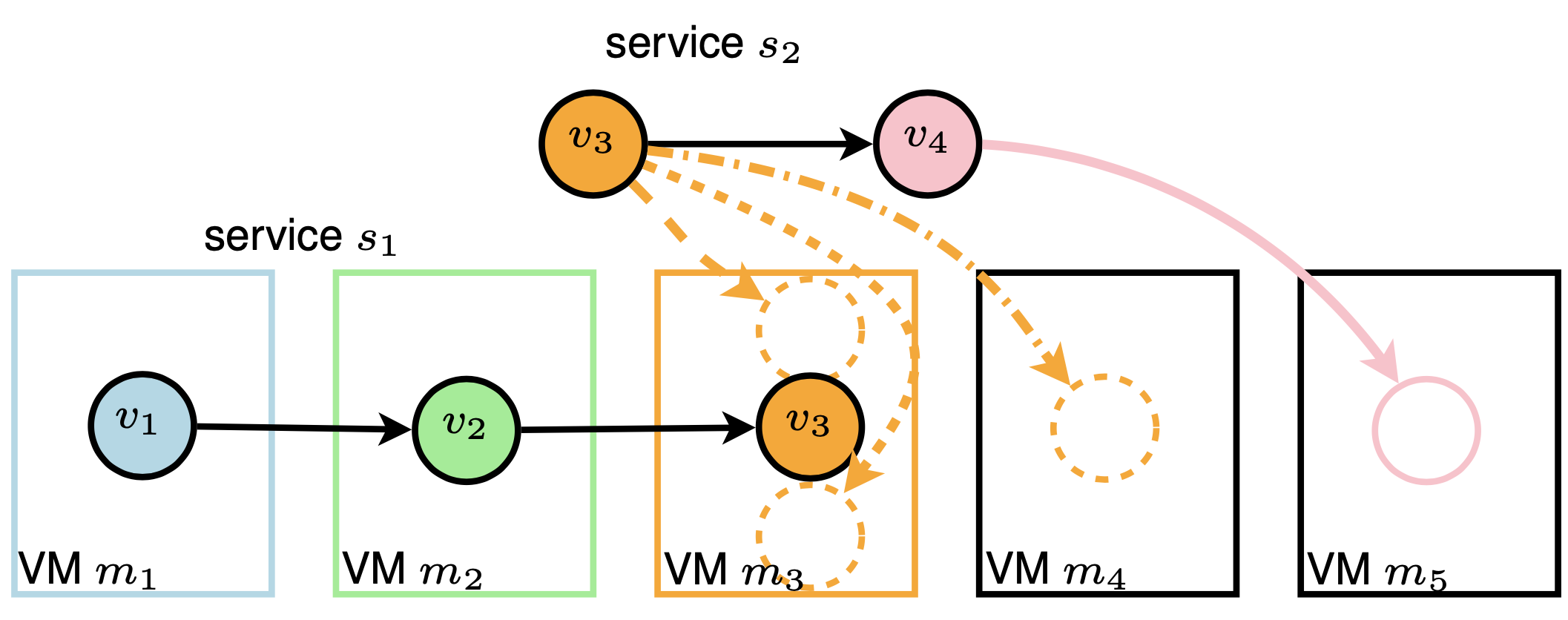}
\caption{
    Example of the VNF-sharing problem. A PoP is serving service~$s_1$, with VNFs~$v_1$--$v_3$ deployed at VMs~$m_1$--$m_3$. It is then requested to deploy~$s_2$, using VNFs~$v_3$ and~$v_4$. No isolation is requested, so services can share VNFs if convenient. For~$v_4$, the only option is devoting an unused VM to it, $m_5$~in the example (pink line). For~$v_3$, instead, there are three options: re-using the instance of~$v_3$ at~$m_3$, giving~$s_2$ a lower priority than~$s_1$ (dashed line); doing the same but giving~$s_2$ a higher priority (dotted line); devoting~$m_4$ -- currently unused -- to~$v_3$ (dash-dotted line), thus having two VMs running~$v_3$.
    \label{fig:idea}
} %caption
\end{figure}

We remark that our work focuses on this {\em VNF sharing} problem, which is different from the one studied in traditional VNF placement studies. 
\Fig{idea} presents a simple instance of the VNF sharing problem: the vertical has requested a new service~$s_2$, and the SDM-X decides at which VM to run each VNF of~$s_2$ (this decision is trivial for~$v_4$, as there is no existing instance of it), and how to prioritize the different services sharing the same VNF.

{\bf Contributions}. 
We make the following main contributions to the VNF-sharing problem:
\begin{itemize}
    \item We observe that allowing {\em flexible} priorities for each VNF and service allows the MNO to meet the KPI targets at a lower cost; 
    \item We present a system model that captures all the relevant aspects of the VNF-sharing problem and the entities it involves, including the capacity-scaling and priority-setting decisions it requires;
    \item leveraging convex optimization, we devise an efficient integrated solution methodology called FlexShare, which is able to make swift, high-quality decisions concerning VM usage, priority assignment, and capability scaling;
    \item we discuss how FlexShare can handle VNF instantiation  and de-instantiation operations;
    \item we formally analyze the computational complexity of FlexShare and its competitive ratio relatively to the optimum;
    \item we study FlexShare's performance using real-world VNF graphs.
\end{itemize}

In the rest of the paper, we begin by motivating the need for flexible priorities across VNFs (\Sec{example}). Then \Sec{model} introduces our system model and problem formulation, while \Sec{algo} describes the FlexShare solution strategy. \Sec{competitive} analyzes FlexShare's performance with respect to the optimal solution. Our reference scenarios and numerical results are discussed in \Sec{results}. Finally, we review related work in \Sec{relwork}, and conclude the paper in \Sec{conclusion}.

\section{The role of priorities}
\label{sec:example}

Before addressing the problem of whether it is convenient to share a VNF among multiple services or not, let us highlight the role of priorities while sharing a VNF instance. Three main approaches can be adopted for VNF sharing:
\begin{itemize}
    \item {\em per-service} priority, associated with each service and constant across different VNFs;
% as in cases~1--3 of \Sec{example};
    \item {\em per-VNF} priority, associated with each service and VNF, thus, given a service, 
it may vary across different VNFs;
%as in the ``flexible priorities'' case in \Sec{example};
    \item {\em per-flow} priority, associated with individual traffic flows (e.g., REST queries) belonging to a service, it may vary across the different VNFs on a per-flow basis.
\end{itemize}
In Example\,1, we focus on the first two steps of the above flexibility ladder, and show how flexible priority assignment can increase the efficiency of handling service flows.

\noindent\rule{1\columnwidth}{0.5mm}
\begin{example}[The importance of flexible priorities]
\label{ex:prios}
Consider the two services, $s_1$ and $s_2$, depicted in \Fig{example}, requested by a vertical specialized in video surveillance systems. 
Recall that services are described as VNF graphs; specifically, 
$s_2$  includes two VNFs executing transcoding and motion detection, respectively, while   $s_1$ is composed of $s_2$ and a VNF performing face recognition. Each VNF should run in its own VM, and assume network transfer times between VMs are negligible.
Adopting a well-established and convenient approach~\cite{placement-infocom,noi-infocom18}, let us model VNFs as M/M/1 traffic queues processing traffic flows, and the services as queuing chains, with arrival rates~$\lambda_1=\SI{2}{flows/\milli\second}$, and~$\lambda_2=\SI{1}{flows/\milli\second}$ respectively.
Also, consider service delay as the main performance metric and let the target average delay be~$D^{\max}_1=D^{\max}_2=\SI{1.1}{\milli\second}$ for both services. Then assume that when given the allocated computation resources,  the service rate of the transcoding and motion detection  is~$\mu_\text{tc}=\mu_\text{md}=\mu=\SI{5}{flows/\milli\second}$, while that of face recognition is $\mu_\text{fr}=\SI{9.15}{flows/\milli\second}$.

To meet the delay targets, $s_2$~flows must traverse the transcoding and motion detection with a combined sojourn time of \SI{1.1}{\milli\second}, while $s_1$~flows must do the same in at most \SI{0.96}{\milli\second} (i.e., the target average delay~$D^{\max}_1$ minus the sojourn time at the face recognition VNF of $\frac{1}{\mu_\text{fr}-\lambda_1}=\SI{0.14}{\milli\second}$). We now show that there is no way of setting {\em per-service} priorities that allow this.

\noindent{\em Case 1: Higher priority to~$s_1$.} This choice would make intuitive sense since $s_1$~flows have to go through more processing stages than $s_2$ flows, within the same deadline. 
In this case, $s_1$~flows incur a sojourn time of~$\frac{1}{\mu-\lambda_1}=\frac{1}{5-2}=\SI{0.33}{\milli\second}$ for each of the common VNFs, resulting in a total delay $D_1=\SI{0.8}{\milli\second}$, well within the target. However, the sojourn time of $s_2$~flows at each of the shared VNFs becomes~\cite[Sec.~3.2]{kleinrock-vol2}:
%\begin{equation}
%\nonumber
$\frac{1/\mu}{\left(1-\frac{\lambda_1}{\mu}\right)\left(1-\frac{\lambda_1+\lambda_2}{\mu}\right)}
=
\frac{1/5}{\left(1-\frac{2}{5}\right)\left(1-\frac{1+2}{5}\right)}\approx\SI{0.83}{\milli\second}$,
%\end{equation}
which results in a total delay of~$D_2\approx\SI{1.66}{\milli\second}>D^{\max}_2$.

\noindent{\em Case 2: Higher priority to~$s_2$.} It is easy to verify that giving higher priority to~$s_2$ implies that $s_1$~misses its target delay.

\noindent{\em Case 3: Equal priority.} Giving the same priority to both services results in a sojourn time of $\frac{1}{\mu-\lambda_1-\lambda_2}=\frac{1}{5-1-2}=\SI{0.5}{\milli\second}$ for each of the common VNFs, and in a total delay of~$D_2=\SI{1}{\milli\second}<D^{\max}_2$ and~$D_1=\SI{1}{\milli\second}+\SI{0.14}{\milli\second}>D^{\max}_1$.

\begin{figure}
\centering
\includegraphics[width=1\columnwidth]{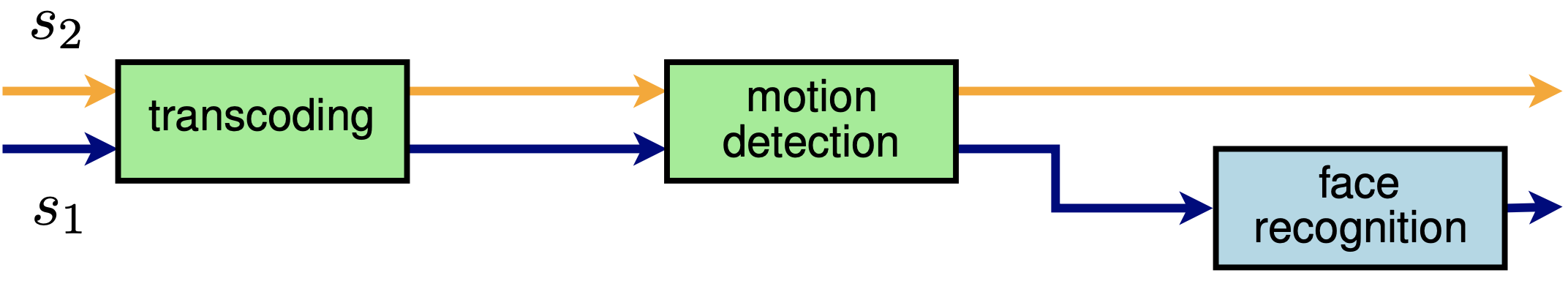}
\caption{\Ex{prios}: two video surveillance services,~$s_1$ and~$s_2$, with $s_2$ including performing transcoding and motion detection VNFs, and $s_1$ composed of $s_2$ and an additional face recognition stage.
    \label{fig:example}
} %caption

\end{figure}

\noindent{\em Flexible priorities.} 
Assume that $s_1$ and $s_2$ have priority in the transcoding and in the motion detection VNF, respectively. Then, $D_1 \mathord{=}\frac{1}{\mu-\lambda_1}\mathord{+}\frac{1/\mu}{\left(1 \mathord{-}\frac{\lambda_2}{\mu}\right)\left(1 \mathord{-}\frac{\lambda_1 \mathord{+}\lambda_2}{\mu}\right)}\mathord{+} 0.14\mathord{=}$
$0.33 \mathord{+} 0.625 \mathord{+} 0.14 \mathord{=}\SI{1.095}{\milli\second}<D^{\max}_1$, and $D_2 \mathord{=}$ $\frac{1/\mu}{\left(1 \mathord{-}\frac{\lambda_1}{\mu}\right)\left(1 \mathord{-}\frac{\lambda_1 \mathord{+}\lambda_2}{\mu}\right)}\mathord{+}\frac{1}{\mu \mathord{-}\lambda_2}$ $\mathord{=}0.83 \mathord{+}0.25 =\SI{1.08}{\milli\second}
<D^{\max}_2$.

\end{example}
\noindent\rule{1\columnwidth}{0.5mm}
In conclusion, the above example shows that relying on merely per-service priorities may result in violation of KPI requirements. In contrast, assigning different per-VNF priorities allows the MNO to meet the vertical's requirements while increasing efficiency in resource usage, hence lowering the costs. 
We later show that per-flow priorities result in even better efficiency.

\section{System model and problem formulation}
\label{sec:model}

\begin{figure}
\centering
\includegraphics[width=.4\textwidth]{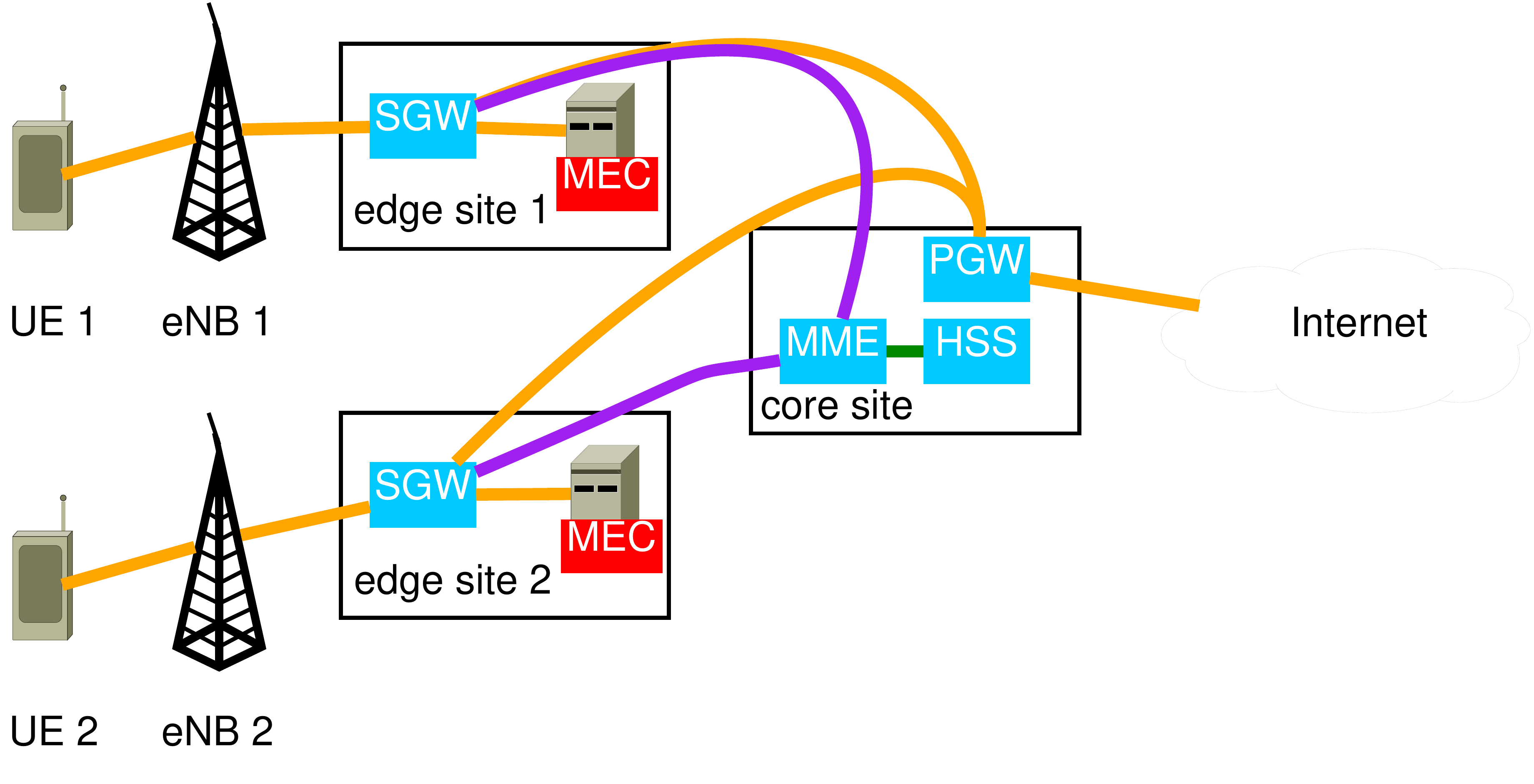}\\
\includegraphics[width=.4\textwidth]{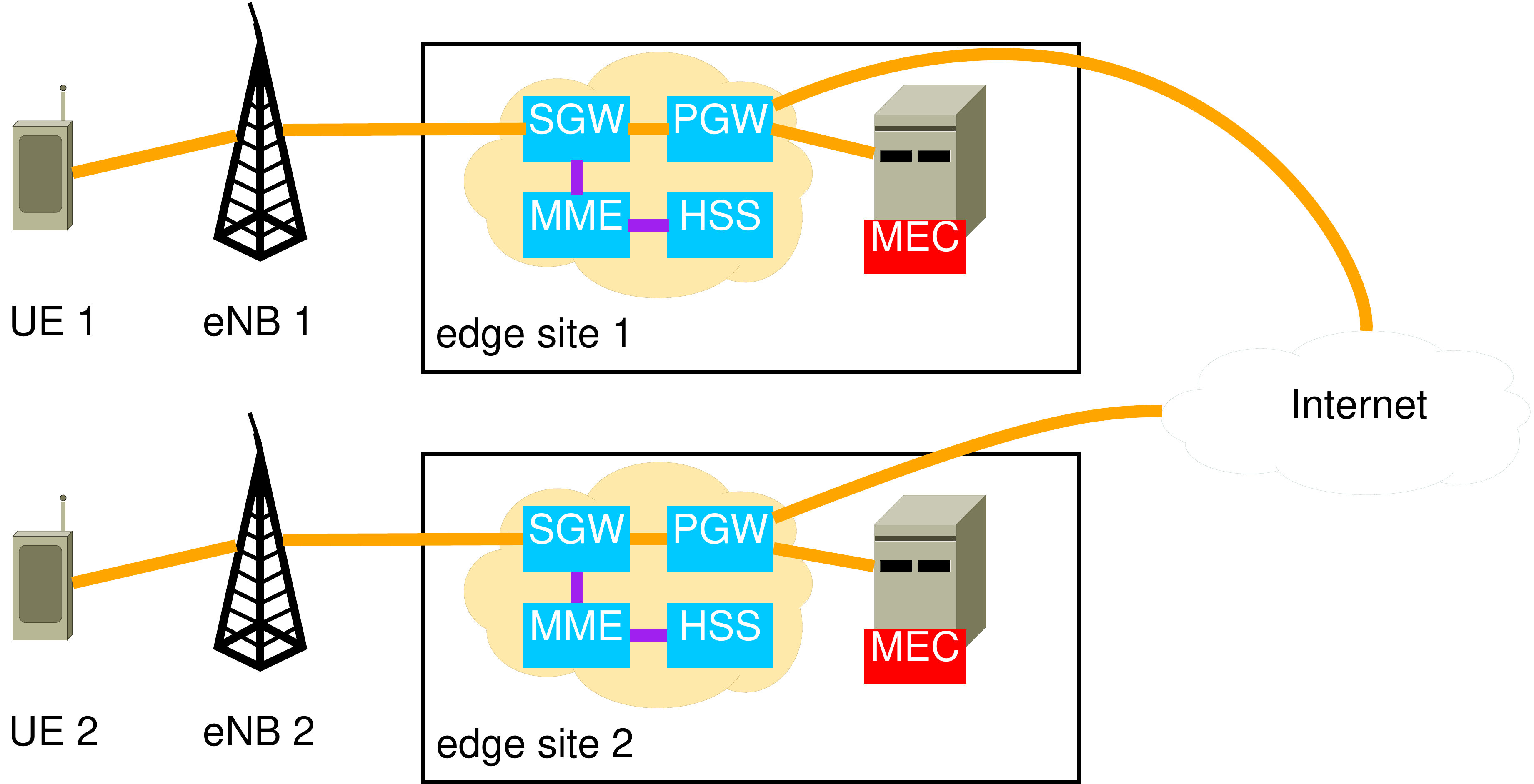}
\caption{
    Two of the possible architectures for MEC-based networks discussed in~\cite{etsi-mec-wp}: distributed EPC (top) and local breakout (bottom). Orange and purple lines correspond to data and control traffic,  respectively.
    \label{fig:mec}
} %caption
\end{figure}

The architecture we consider reflects real-world deployments of MEC-based networks, as described in the ETSI specifications~\cite{etsi-mec-wp}. Possible deployments differ in aspects like the location of the EPC, but all include multiple {\em edge sites}, i.e., PoPs hosting the MEC servers and some (as in \Fig{mec}(top)), or all (as in \Fig{mec}(bottom)), of the EPC network functions.

As mentioned, the network and computing resources available in the MEC in \Fig{mec} have to be properly orchestrated and managed. In particular, referring to the 5G architecture in \cite{5gppp-architecture},   the SDM-X has to use the VMs under its control to provide the newly-requested services with the required KPIs and at the minimum cost. It should thus make decisions on
\begin{inparaenum}[(i)]
\item {\em whether} existing VNF instances should to be shared;
\item if so, {\em how} to assign the priorities to different services sharing the same VNF instance; and last
\item {\em scaling} the computational capabilities of the VMs if needed and possible.
\end{inparaenum}

We choose to focus on single PoP decisions and account for the system overheads, which, in general, are given by network delays and processing times.  However, a study of the ongoing work in the datacenter networking community (see, e.g.,~\cite{datacenter-survey}) suggests that switching is highly optimized, hence network delays are often minimal. Also, network topologies tend to be highly regular, hence delays are similar across any pair of VMs. Therefore, it is fair to neglect the network delays {\em within a single PoP}, and consider the processing times only.

Below, we describe the model and the entities that are involved in the VNF-sharing problem (\Sec{sub-model}), followed by defining the problem objective and constraints (\Sec{problem}).

\subsection{System model}
\label{sec:sub-model}

The system model includes VMs~$m\in\Mc$, and VNFs~$v\in\Vc$~\footnote{VNFs can, in general, include multiple virtual deployment units (VDUs); without loss of generality,  we assume that each VNF includes only one VDU. Also, we assume that no VNF  requires isolation.}.
Each VM runs (at most) one VNF, modeled as an M/M/1 queue with FIFO queuing and preemption, as widely assumed in recent works~\cite{placement-infocom,noi-infocom18,DBhamare17}. Also, let $C(m)$ be the maximum computation capability to which VM $m$ can be scaled up. We underline that, although in this work we focus on computational capability, we could adjust our model to focus on memory and storage as well.  We refer to a VM as active if it hosts a VNF, and we express through binary variables~$y(v,m)$ whether VM~$m$ runs VNF~$v$. 

VNFs vary in computational requirements, which are modeled through parameter~$l(v)$, expressing how many units of computational capability are needed to process one flow for VNF $v$ per time unit.
For example, a VNF with requirement~$l(v)=1$ running on a VM~$m$ with capability~$\mu(m)=1$ takes $l(v)/\mu(m)=1$~time unit to process a flow. Using the same VM for a VNF with requirement~$l(v)=2$ yields a processing time of $l(v)/\mu(m)=2$~time units per flow. Note that $l(v)$ values do not depend on the actual service using VNF~$v$, but on~$v$ only.

Services~$s\in\Sc$ include one or more VNFs, and flows belonging to service~$s$ arrive at VNF~$v$ with a rate~$\lambda(s,v)$; VNFs that are not used by a certain service have $\lambda$-values equal to~$0$. 
Through the~$\lambda(s,v)$ parameters, we can account for arbitrarily complex service (VNF) graphs where the number of flows can change between VNFs, and some flows may visit the same VNF more than once.
For sake of simplicity, in this paper we focus on the maximum average delay~$D^{\max}(s)$ of service~$s$ as the target KPI\footnote{Note that our model can be extended to additional KPIs.}.

Each VM uses a quantity~$\mu(m)\le C(m)$ of computational capability that can be dynamically adjusted.  Given $\mu(m)$,  VNF~$v$ deployed at VM~$m$ processes flows at rate~$\frac{\mu(m)}{l(v)}$. Finally, binary variables~$x(s,v,m)$ express whether service~$s$ uses the instance of VNF~$v$  at VM~$m$; this allows us to model the assignment of distinct services requiring the same VNF to different VMs that run the VNF. For clarity, we summarize the above parameters and variables in~\Tab{notation}.

\begin{table}
\caption{Notation ($\dagger$~denotes variables of the modified problem described in \Sec{sub-assignment})
    \label{tab:notation}
} %caption
\scriptsize
\begin{tabularx}{\columnwidth}{|l|l|X|}
\hline
{\bf Symbol} & {\bf Type} & {\bf Meaning} \\
\hline
$\Mc=\{m\}$ & Set & Set of VMs \\
\hline
$\Vc=\{v\}$ & Set & Set of VNFs \\
\hline
$\Sc=\{s\}$ & Set & Set of services \\
\hline
$C(m)$ & Parameter & Maximum capability to which VM~$m$ can be scaled up \\
\hline
$l(v)$ & Parameter & Computational capability needed to process one flow unit for VNF~$v$ \\
\hline
$\lambda(s,v)$ & Parameter & Arrival rate of flows of service~$s$ for VNF~$v$ \\
\hline
$D^{\max}(s)$ & Parameter & Target delay for service~$s$ \\
\hline
$\kappa_f(m)$ & Parameter & Fixed cost incurred when activating VM~$m$ \\
\hline
$\kappa_p(m)$ & Parameter & Proportional cost incurred when using one unit of computational capability for VM~$m$ \\
\hline
$p(s,v)$ & Parameter & Per-VNF priority of service~$s$ at VNF~$v$ \\
\hline
%$j$ & Parameter & Variation applied when assigning per-flow priorities \\
%\hline
$\mu(m)$ & Decision variable & Computational capability to use for VM~$m$ \\
\hline
$y(v,m)$ & Decision variable & Whether VM~$m$ runs VNF~$v$ \\
\hline
$x(s,v,m)$ & Decision variable & Whether flows of service~$s$ use the instance of VNF~$v$ running at VM~$m$ \\
\hline
\rule{0pt}{7pt} \hspace{-1.5mm} $\tilde{\Lambda}(s,v)$ & \hspace{-1mm} Decision variable.$^\dagger$ & \hspace{-1mm}  Arrival rate of flows for VNF~$v$ on the same VM as the one servicing $s$, that are given priority over flows of service~$s$ \\
\hline
$S(s,v)$ & Auxiliary decision variable & Sojourn time of flows of service~$s$ for VNF~$v$ \\
\hline
$\Lambda(s,v)$ & Auxiliary decision variable & Arrival rate of flows for VNF~$v$ on the same VM as the one servicing $s$, that are given priority over flows of service~$s$ \\
\hline
$\pi(s,v)$ & Random variable &
Describes the priority assigned to flows of service~$s$ upon entering VNF~$v$\\
\hline
\end{tabularx}
\end{table}

\subsection{Problem formulation}
\label{sec:problem}

We now discuss the objective of the VNF-sharing problem and the constraints we need to satisfy.

{\bf Objective.}
The high-level goal of the MNO is to minimize its incurred cost, which consists of two components: a fixed cost,~$\kappa_f(m)$, paid if  VM~$m$ is activated, and a proportional cost,~$\kappa_p(m)$, paid for each unit of computational capability used therein. The objective is then given by:
\begin{equation}
\label{eq:obj}
\min_{y,\mu}\sum_{m\in\Mc}   \left(  \kappa_f(m)  \sum_{v\in\Vc} y(v,m) +\kappa_p(m)\mu(m)\right).
\end{equation}

{\bf VM capability and VNF instances.}
We must account for the 
maximum value~$C(m)$ to which the capability~$\mu(m)$ of each VM~$m$ can be scaled up: 
\begin{equation}
\label{eq:capacity}
\mu(m)\leq C(m)\quad \forall m\in\Mc.
\end{equation}
Also, at most one VNF can run in any single VM:
\begin{equation}
\label{eq:y}
\sum_{v\in\Vc}y(v,m)\leq 1,\quad\forall m\in\Mc,
\end{equation}
and only active VMs can be used for handling flows, i.e.,
\begin{comment}
%%% removed this because we are not saying that we allow only one instance of each VNF for each service
\begin{equation}
\label{eq:one-instance-service}
\sum_{m\in\Mc,v\in\Vc}x(s,v,m)\leq 1,\quad\forall s\in\Sc;
\end{equation}
\end{comment}
\begin{equation}
\label{eq:only-existing}
y(v,m)\geq x(s,v,m),\quad\forall s\in\Sc,v\in\Vc, m\in\Mc.
\end{equation}

{\bf Service times.}
Each service~$s$ has a maximum average service time~$D^{\max}(s)$ that must be maintained. Since we assume that processing time is the dominant component of service time, this is equivalent to imposing:
\begin{equation}
\label{eq:max-service-time}
\sum_{v\in\Vc}S(s,v) \leq D^{\max}(s),\quad\forall s\in\Sc,
\end{equation}
where~$S(s,v)$ is the {\em sojourn time} (i.e., the time spent waiting or being served) experienced by flows of service~$s$ for VNF~$v$. By convention, we set~$S(s,v)=0$ if service~$s$ does not require VNF~$v$.

As detailed below, sojourn times, in turn, depend on:
\begin{itemize}
    \item the computational capability~$l(v)$ required for handling any flow for a VNF;
    \item the traffic flow arrival rate at the VNFs~$\lambda(s,v)$;
    \item the priority of the traffic flows at the traversed VNFs (to be detailed in the sequel);
    \item the computational capability~$\mu(m)$ assigned to the VM hosting the VNF instance processing the flow.
\end{itemize}
Using~\cite[Sec.~3.2]{kleinrock-vol2} and~\cite{dispensa}, we can generalize the expression used in \Ex{prios} and write the sojourn time of flows of service~$s$ for VNF~$v$ as:
\begin{equation}
\label{eq:S}
S(s,v){=}\frac{l(v)}{\mu(\bar{m})}\frac{1}{1-l(v)\frac{\Lambda(s,v)}{\mu(\bar{m})}}\frac{1}{1-l(v)\frac{\Lambda(s,v)+\lambda(s,v)}{\mu(\bar{m})}},
\end{equation}
where~$\bar{m}$ is the VM hosting the instance of VNF~$v$ used by service~$s$, i.e., where $x(s,v,\bar{m})=1$.

In \Eq{S}, $\Lambda(s,v)$ represents the arrival rate of flows (of any service) for the instance of VNF~$v$ hosted on $\bar{m}$, that are given a priority higher than a generic flow of service~$s$ for VNF~$v$.
Let~$\pi(s,v)$ be the random variable describing the priority assigned to flows of service~$s$ at VNF~$v$, then:
\begin{equation}
\label{eq:def-lambda}
\Lambda(s,v)=\sum_{t\in\Sc}\PP\left(\pi(t,v)>\pi(s,v)\right)\lambda(t,v).
\end{equation}
The intuitive meaning of \Eq{def-lambda} is that~$\Lambda(s,v)$ grows as it becomes more likely that flows of other services~$t\neq s$ are given higher priority over flows of service~$s$.

The expression of~$\Lambda(s,v)$ depends on the type of the $\pi(s,v)$~variables. In App.~\ref{app:computelambda}, we show how $\Lambda(s,v)$~values can be computed for the two priority assignments discussed in \Sec{example}, i.e., per-VNF priorities and uniform, per-flow priorities.

{\bf Problem complexity.} 
In the general case, the expression of~$\Lambda(s,v)$ is not guaranteed to be linear, convex, or even continuous. It follows that no hypothesis can be made about the complexity of the problem of setting the priorities so as to optimize \Eq{obj}: solving such a problem may require to search over all possible distributions of~$\pi(s,v)$, which would be prohibitively complex even in small-scale scenarios.

Indeed, it is possible to reduce any instance of the bin-packing problem, which is NP-hard, to a {\em simplified} instance of our problem where (i) there is only one type of VNF, (ii) all services have the same target KPIs, and (iii) $\kappa_p=0$ for all VMs. Specifically:
\begin{itemize}
    \item VMs correspond to bins;
    \item the capability of VMs correspond to the size of the capacity of the bin they are associated with;
    \item services correspond to items;
    \item traffic flow arrival rates~$\lambda(s,v)$ correspond to the size of items.
\end{itemize}

{\bf Non-negligible network delays.}
As mentioned earlier, network delay within individual PoPs are very small compared to processing times, hence, our model neglects them. However, it is worth stressing that our model can be extended to account for arbitrary-delay scenarios, by introducing the following new parameters:
\begin{itemize}
    \item for each pair~$(m_1,m_2)$ of VMs in~$\Mc$, a network delay~$d(m_1,m_2)$;
    \item for each pair~$(v_1,v_2)$ of VNFs in~$\Vc$, a parameter~$\rho(v_1,v_2)\in[0,1]$ expressing the fraction of traffic that visits~$v_2$ immediately after visiting~$v_1$.
\end{itemize}
Given the above parameters, the total network delay is:
\begin{equation}
\label{eq:netdelay}
\sum_{m_1,m_2\in\Mc}\sum_{v_1,v_2\in\Vc} d(m_1,m_2) y(v_1,m_1)y(v_2,m_2)\rho(v_1,v_2).
\end{equation}
Eq.\,\Eq{netdelay} can be read as follows: a network delay~$d(m_1,m_2)$ is incurred when (i) a VNF~$v_1$ is deployed at~$m_1$, and (ii) the subsequent VNF~$v_2$ is deployed at~$m_2$. The quantity in \Eq{netdelay} should then be added to the first member  of the delay constraint \Eq{max-service-time}, thereby ensuring that the combination of processing and network delays is below the delay target~$D^{\max}$.

\section{The FlexShare solution strategy}
\label{sec:algo}

\begin{figure}
\centering
\includegraphics[width=1\columnwidth]{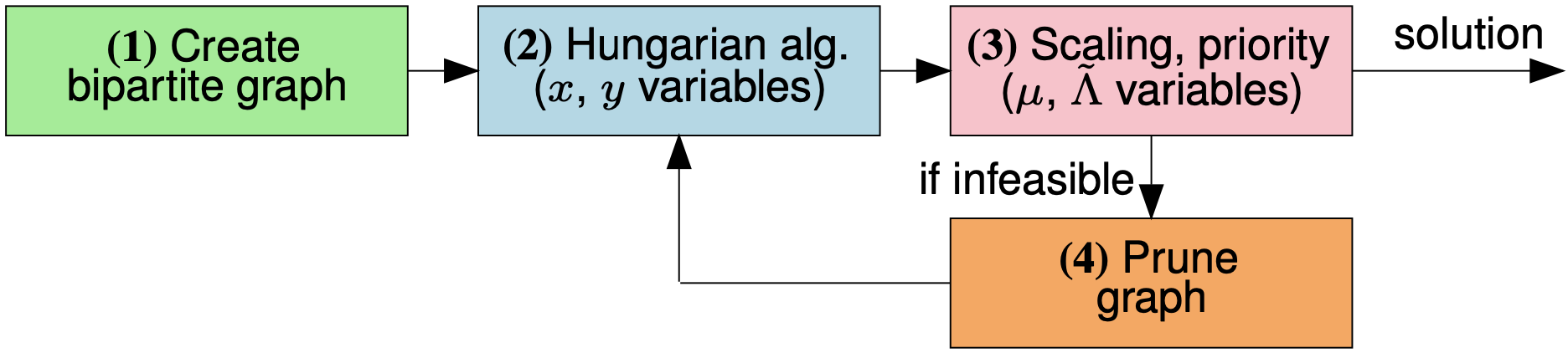}
\caption{
    The FlexShare strategy. Step~1 builds a bipartite graph showing which VMs {\em could} run each VNF. Step~2 runs the Hungarian algorithm on such a graph to obtain the optimal values for the~$x$- and~$y$-variables. Step~3 solves a convex variant of the original problem in \Sec{problem}. If feasible, its variables ($\mu$, $\tilde{\Lambda}$) are used to determine the scaling and the priorities; otherwise, the bipartite graph is {\em pruned} (step~4) and the procedure restarts from step~2.
    \label{fig:flow}
} %caption

\end{figure}

In light of the problem complexity, we propose a fast, yet highly effective solution strategy named FlexShare, which runs iteratively and consists of four main steps as outlined in \Fig{flow}.
The algorithm considers services one by one, adjusting its solution for every new service being deployed.
The first step, detailed in \Sec{sub-bipartite}, consists of building a bipartite graph including the VNFs of the new service that are to be deployed, and the VMs that are active or can be activated. 
The edges of the graph express the possibility of using a VM to provide a VNF, either by sharing an existing instance of the VNF or by deploying a new one. Edges are labeled with the cost associated with each decision, i.e., the~$\kappa_p$ and/or~$\kappa_f$ terms contributing to the objective \Eq{obj}.
In step~2, also described in \Sec{sub-bipartite}, we use the Hungarian algorithm~\cite{hungarian} on the generated bipartite graph to get the optimal minimum-cost assignment of VNFs to VMs, i.e., the~$x$- and $y$-variables.

Given these decisions, step~3 aims at assigning the priorities and finding the amount of computational capability to use in every VM. To this end, a
simpler (namely, {\em convex}) variant of the problem defined in \Sec{problem} is formulated and solved, as detailed in \Sec{sub-assignment}.

If step~3 results in an infeasible problem, we {\em prune} the bipartite graph (step~4). The underlying intuition is that a cause for infeasibility is overly aggressive sharing of existing VNF instances. Therefore, as detailed in \Sec{sub-prune}, we prune from the bipartite graph edges that result in an overload of VMs. After pruning, the algorithm starts a new iteration with step~2. Moving from one iteration to the next means reducing the likelihood that VNF instances are shared between services, and thus increasing the cost incurred by the MNO, due to the~$\kappa_f$ fixed cost terms. The procedure stops as soon as it finds a feasible solution. Without loss of generality, we present FlexShare in the case where there are sufficient resources, i.e., enough VMs, to deploy all the requested services.

\subsection{Steps 1--2: Bipartite graph and Hungarian algorithm}
\label{sec:sub-bipartite}

{\bf The bipartite graph.}
The bipartite graph represents
\begin{inparaenum}[(i)]
\item the possible VNF assignment decisions, i.e., which VNFs can be provided at which VMs and which VNFs can be shared among services, and
\item the associated cost incurred by the MNO.
\end{inparaenum}

More formally, the bipartite graph is created according to the following rules:
\begin{enumerate}
    \item a vertex is created for each VNF and for each VM;
    \item an edge is drawn from every VNF to every unused VM;
    \item an edge is drawn from every VNF to every VM currently running the same VNF, provided that the maximum computational capability of the VM is sufficient to guarantee stability.
\end{enumerate} 

Denote by~$\bar{s}$ the service now being deployed. For every VNF~$v$ required by~$\bar{s}$, and every VM~$m$ either running $v$ or not yet activated, $m$ can satisfy the request of $\bar{s}$ for $v$ while ensuring stability if
\begin{equation}
\label{eq:check-stability}
l(v) \sum_{s\in\Sc}
\left [
\left(x(s,v,m)+\ind{s=\bar{s}}\right)\lambda(s,v)
\right ] < C(m),
\end{equation}
i.e., if the total load on VM~$m$ is no larger than its maximum capability~$C(m)$.
% The first member of \Eq{check-stability} is the total load imposed on VM~$m$ by services that are already being served therein, if any, {\em and} the new one~$\bar{s}$ (for which the indicator function is one). Through \Eq{check-stability}, we can then check that this quantity is lower than the maximum VM capability~$C(m)$.

\begin{figure}
\centering
\includegraphics[width=.25\textwidth]{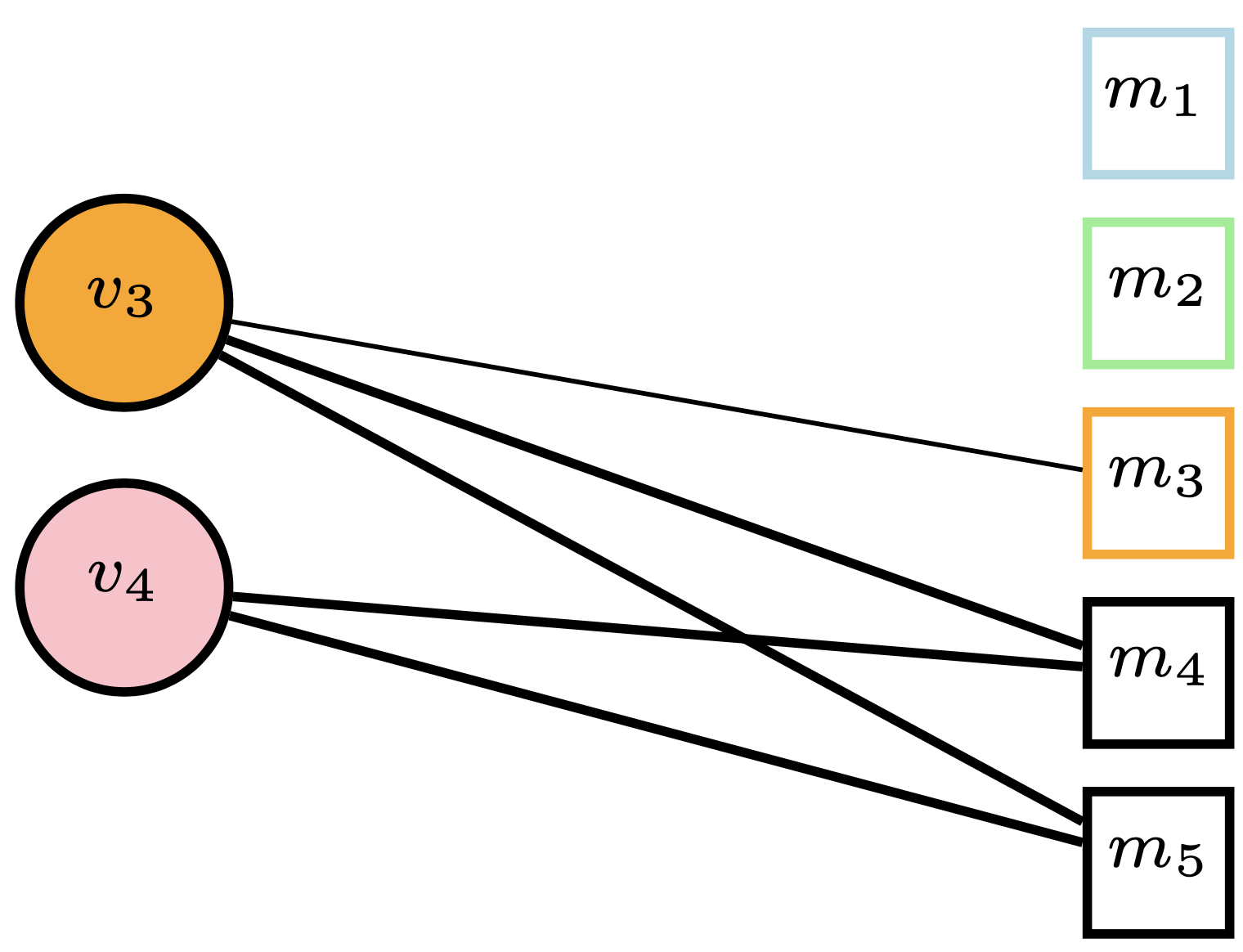}
\caption{
    The bipartite graph generated when trying to deploy service~$s_2$, as shown in \Fig{idea}. Graph vertices correspond to VNFs in~$s_2$ (left) and VMs (right); edges represent the possible assignment decisions. Edges connecting currently-unused VMs, i.e.,~$v_4$ and~$v_5$, are thicker because they are associated with a higher cost (due to the~$\kappa_f$ component).
    \label{fig:bipartite}
} %caption
\end{figure}

Note that \Eq{check-stability} does {\em not} imply that~$\bar{s}$, or any other existing services, can be served {\em in time}, i.e., while satisfying their delay constraints; indeed, this depends on the priority and computational capability assignment decisions, and cannot be checked at the graph generation time. The purpose of step~1 is to generate a graph accounting for all possible assignment options, that may {\em potentially} result in a feasible solution.
\Fig{bipartite} provides an example of a bipartite graph, representing the options available in the scenario depicted in \Fig{idea}: VNF~$v_3$ can be provided at VMs~$m_3$ (which already runs~$v_3$), or at $m_4$ or $m_5$ (which are both currently yet inactive); VNF~$v_4$ can only be run on either~$m_4$ or~$m_5$.

The cost of each edge connecting VM~$m$ with VNF~$v$ is given by the following expression:
\begin{equation}
\label{eq:edge-cost}
\left(1-\sum_{v}y(v,m)\right)\kappa_f(m)+\kappa_p(m)\left( l(v)\lambda(\bar{s},v)+\epsilon\right).
\end{equation}
In \Eq{edge-cost}, the first term is the fixed cost associated with activating VM~$m$, which is incurred only if $m$~is not already active (the summation can be at most~$1$, as per \Eq{y}). The second term is the proportional cost associated with the additional computation capability needed at VM~$m$ to guarantee stability, with~$\epsilon$ being a positive,  arbitrarily small value.

{\bf Hungarian algorithm and assignment decisions.}
The Hungarian algorithm~\cite{hungarian} is a combinatorial optimization algorithm with polynomial (cubic) time complexity in the number of edges in the graph. When applied to the bipartite graph we generate, it selects a subset of edges such that
\begin{inparaenum}[(i)]
\item each VNF is connected to exactly one VM, and
\item the total cost of the selected edges is minimized.
\end{inparaenum} 

Selected edges map to assignment decisions. Specifically, for each selected edge connecting VNF~$v$ and VM~$m$, we set $y(v,m)\gets 1$ and~$x(\bar{s},v,m)\gets 1$, i.e., we activate~$m$ (if not already active)  deploying therein an instance of~$v$,  and use it to serve service~$\bar{s}$. The obtained values for the~$x$ and $y$-variables are used in step~3 to decide priorities and computational capability assignment, as set out next.

\subsection{Step~3: Priority and scaling decisions}
\label{sec:sub-assignment}

The purpose of step~3 of the FlexShare procedure is to decide the priorities to assign to each VNF and service, as well as any needed scaling of VM computation capability. Since the complexity of the problem stated in \Sec{problem} depends on the
presence of the~$\pi(s,v)$ variables, we proceed as follows:
\begin{enumerate}
    \item we formulate a {\em simplified} problem, which contains no random variables and is guaranteed to be convex;
    \item we use the variables of the simplified problem to set the~$\mu(m)$ variables of the original problem, as well as the parameters of the distribution of the~$\pi(s,v)$ variables.
\end{enumerate}

{\bf Convex formulation.}
To avoid dealing with probability distributions, we replace the $\Lambda(s,v)$~auxiliary variables of the original problem with independent variables~$\tilde{\Lambda}(s,v)$, thus dispensing with \Eq{def-lambda}. Given $x$ and $y$, the decision variables of the modified problem are~$\tilde{\Lambda}(s,v)$ and~$\mu(m)$, while the objective is still given by \Eq{obj}. Having~$\tilde{\Lambda}(s,v)$ as a variable means deciding (intuitively) how many higher-priority traffic flows each incoming flow will find. Such values are later mapped to the parameters of the distributions of~$\pi(s,v)$.

If we solve the modified problem with no further changes, the optimal solution would always yield~$\tilde{\Lambda}(s,v)=0,\forall s,v$, i.e., no flow ever encounters higher-priority ones, which is clearly not realistic. To avoid that, we mandate that the average behavior, i.e., the average number of higher-priority flows met, is the same as in the original problem:
\begin{equation}
\label{eq:minlambda}
\sum_{s\in\Sc}\tilde{\Lambda}(s,v)=\frac{|\Sc|}{2}\sum_{s\in\Sc}\lambda(s,v),\quad\forall v\in\Vc.
\end{equation}
The intuition behind \Eq{minlambda} is that each $\Lambda(s,v)$-value (in the original problem) is the sum of several $\lambda$-values, i.e., the services arrival rates. The $\lambda$-value associated with the highest-priority service will contribute to~$|\Sc|-1$ $\Lambda(s,v)$-values, the one associated with the second-highest-priority service will contribute to $|\Sc|-2$ $\Lambda(s,v)$-values, and so on. On average, each $\lambda$-value contributes to~$\frac{|\Sc|}{2}$ $\Lambda(s,v)$-values. Finally, recall that~$\lambda(s,v)=0$ if service~$s$ does not use VNF~$v$.

It  can be proved that the modified problem is convex: 
\begin{property}
\label{prop:convex}
The problem of minimizing \Eq{obj} subject to constraints \Eq{capacity}--\Eq{max-service-time} and \Eq{minlambda}, is convex.
\end{property}
\begin{IEEEproof}
For the problem to be convex, the objective and all constraints must be so. Our expressions are linear, and thus convex. However, \Eq{max-service-time} contains $S(s,v)$-terms, which have to be proven to be convex. We do so by computing the second derivative of the expression~$S(s,v)$ in the~$\mu(m)$ and~$\Lambda(s,v)$ variables. It is easy to verify that, since the quantities $\tilde{\Lambda}(s,v)$, $\mu(m)$, $\lambda(s,v)$ and  $l(v)$ are all positive and  the system is stable (i.e.,  $l(v)\Lambda(s,v)<\mu(m)$ and~$l(v)(\Lambda(s,v)+\lambda(s,v))<\mu(m)$),  both derivatives are positive, which proves the thesis.
\end{IEEEproof}
Note that the above property implies that the modified problem is  solvable in polynomial time
(in the problem size, which depends on the number of VNFs, VMs, and already-deployed services)
through off-the-self, commercial solvers.

{\bf Setting the variables of the original problem.}
After solving the convex problem described above, we can use the optimal solution thereof to make scaling decisions, i.e., to set the~$\mu(m)$ variables in the original problem, as well as the priorities, i.e., the parameters of the distribution of~$\pi(s,v)$. For $\mu(m)$, we can simply use the corresponding variables in the simplified problem, which have the same meaning and are subject to the same constraints.

As far as priorities are concerned, the procedure to follow depends on the priority assignment adopted in the system at hand, hence, on the type of the variables~$\pi(s,v)$. With reference to the per-VNF and per-flow priority assignments used in \Sec{example} and to the computations performed in Appendix~\ref{app:computelambda}, the following holds:
\begin{itemize}
    \item when per-VNF priorities are used, we set the $p(s,v)$~values in Appendix~\ref{app:computelambda} in such a way that services associated with a higher~$\Lambda(s,v)$ have lower priority, e.g., by imposing that $p(s,v)\gets-\tilde{\Lambda}(s,v)$;
    \item when per-flow priorities are supported, then we can solve a system of linear equations where the~$\tilde{\Lambda}(s,v)$ from the solution of the simplified problem are known terms, the $r(s,v)$~quantities are the unknowns, and equations have the form of \Eq{def-q} and \Eq{def-lambda-rlevel} in Appendix~\ref{app:computelambda}.
\end{itemize}

Regardless the way priorities are assigned, it is important to stress that our approach has general validity and can be combined with {\em any type} of priority distribution.

\subsection{Step~4: graph pruning}
\label{sec:sub-prune}

If the problem we solve in step~3 (priority and scaling decisions) is infeasible, a possible cause lies in the decisions made in step~2, i.e., the~$x$ and $y$ variables. Thus, we restart from step~2  considering a different bipartite graph, more likely to result in a feasible problem.

To this end, we consider the {\em irreducible infeasible set}~\cite{iis} (IIS) of the problem instance solved in step 3, i.e., the set of constraints therein that, if removed, would yield a feasible problem. Given the IIS, we proceed as follows:
\begin{enumerate}
    \item we identify constraints in the IIS of type \Eq{capacity}, thus, a set of VMs that would need more capability;
    \item among such VMs, we select those that are used by the newly-deployed service~$\bar{s}$;
    \item among them, we identify the one that is the closest to instability, i.e., the VM~$m^\star$ minimizing the quantity~$C(m)-\sum_{s\in\Sc} x(s,v^\star,m)l(v^\star)\lambda(s,v^\star)$, where $v^\star$~is the VNF deployed at~$m$;
    \item we prune from the bipartite graph the edge $(v^\star$,$m^\star)$.
\end{enumerate}
The intuitive reason for this procedure is that a cause for delay constraints violations is that the newly-deployed service~$\bar{s}$ is causing one of the VMs it uses to operate too close to instability, and thus with high delays. By removing the corresponding edge from the bipartite graph, we ensure that VM~$m^\star$ is not used by service~$\bar{s}$.

Note that we are guaranteed that the IIS contains at least one constraint of type \Eq{capacity} thanks to the following result:
\begin{theorem}
\label{thm:iis}
Every infeasible instance of the modified problem presented in \Sec{sub-assignment} includes at least one constraint of type \Eq{capacity} in its IIS.
\end{theorem}
\begin{IEEEproof}
The constraints of the modified problem are of type \Eq{capacity}--\Eq{max-service-time} and \Eq{minlambda}. Proving that there is a constraint of type \Eq{capacity} in the IIS is equivalent to proving that we can solve a violation of the other types of constraint by violating one or more constraints of type \Eq{capacity}.
Indeed, if a max-delay constraint of type \Eq{max-service-time} is violated, we can make the capacity of the VNF used by that service arbitrarily high; so doing, we can solve the violation of \Eq{max-service-time} at the cost of  violating \Eq{capacity}. 
Similarly, solving a violation of \Eq{minlambda} requires increasing the $\tilde{\Lambda}$-values, which in turn increases the sojourn times and results in a violation of \Eq{max-service-time}-type constraints, thus reducing to the previous case.
\end{IEEEproof}
FlexShare then restarts with step~2, where the Hungarian algorithm takes as an input the pruned bipartite graph.

It is worth stressing that the choice of the edge to prune from the bipartite graph only depends upon the quantity~$C(m)-\sum_{s\in\Sc} x(s,v^\star,m)l(v^\star)\lambda(s,v^\star)$, as specified in item~3 above. Therefore, such a decision is independent on the order in which VNFs appear in the VNF graph of the service at hand.

\subsection{Computational complexity}
\label{sec:polynomial}

The FlexShare strategy has polynomial worst-case computational complexity. Specifically:
\begin{itemize}
    \item step~1 involves a simple check over at most~$|\Vc||\Mc|$ VNF/VM pairs;
    \item step~2, the Hungarian algorithm, has cubic complexity in the number of nodes in the graph~\cite{hungarian};
    \item step~3 requires solving a convex problem, as proven in \Prop{convex}, and the resulting complexity is also cubic;
    \item step~4 iterates over at most~$|\Mc|$ constraints of type \Eq{capacity}, and thus it has linear complexity;
    \item the whole procedure is repeated for (at most) as many times as there are edges in the original bipartite graph.
\end{itemize}

\subsection{Managing service de-instantiations}
\label{sec:drop}

So far we have described how FlexShare deals with requests to instantiate new services. In real-world scenarios, services  have a finite lifetime, hence, they will have to be {\em de-instantiated} as such a lifetime expires. This can lead to suboptimal situations, such as the one  in \Ex{drop}.

\noindent\rule{1\columnwidth}{0.5mm}
\begin{example}[The effect of de-instantiating services]
\label{ex:drop}

Consider three services~$s_1,\dots ,s_3$, all including VNF~$v_1$ and all having~$\lambda(s,v_1)=1$. Also assume that~$l(v_1)=1$ and that, in order to meet their deadlines, all services need that the service time at~$v_1$ be lower than~$S^{\max}(s,v_1)=\SI{1}{\milli\second}$. Finally, assume that there are two available VMs, $m_1,m_2$, both having maximum capability~$C(m)=5$.

Upon receiving the request for~$s_1$, FlexShare will create an instance of~$v_1$ in VM~$m_1$, resulting in a service time of~$S(s_1,v_1)=\SI{0.25}{\milli\second}$. The same VNF instance can be used for~$s_2$, which would get -- assuming, without loss of generality, that it is assigned a lower priority\footnote{In this example, all services have the same flow arrival rate and the same maximum delay, hence, priorities do not influence whether a certain placement is feasible or not.} -- a service time of~$S(s_2,v_1)=\SI{0.42}{\milli\second}$. As for~$s_3$, re-using the instance of~$v_1$ at~$m_1$ would result in a service time~$S(s_3,v_1)=\SI{1.04}{\milli\second}$, exceeding the target delay; therefore, a new instance of~$v_1$ is created at~$m_2$.

After its lifetime expires, service~$s_2$ is de-instantiated, and we are left with two VMs, $m_1$ and~$m_2$, both running instances of~$v_1$. This is a suboptimal situation, since~$s_3$ could now use the instance of~$v_1$ at~$m_1$, without the need to keep~$m_2$ active. 
\end{example}
\noindent\rule{1\columnwidth}{0.5mm}
Situations like the one in \Ex{drop} can happen whenever services have limited lifetimes, and cannot be avoided {\em a priori}. However, they can be effectively managed {\em a posteriori}. 
Specifically, we can periodically check whether there are two VMs, $m_1,m_2$, and a VNF~$v$ such that (i) both VMs run instances of~$v$, and (ii) there exists a priority and capability assignment within~$m_1$ such that all the services currently using the~$v$ instance in~$m_2$ can use the instance in~$m_1$ instead, while experiencing the same service times. It is easy to see that such a check can be performed by solving a reduced version of the problem presented in \Sec{sub-assignment}, i.e., in polynomial time. 
If the check  indicates that services ought to be moved, then FlexShare indeed moves all services currently using~$m_2$ (i.e.,~$x(s,v,m_2)=1$) to~$m_1$ (i.e.,~$x(s,v,m_1)\gets 1$) and deactivates $m_2$ (i.e.,~$y(m_2,v)\gets 0$).
In such a case we say that the VNF instance deployed in $m_1$ and that deployed in $m_2$ were {\em merged} into $m_1$. 
We recurrently execute this procedure as long as it reduces the overall cost, 
in particular,  prior to processing any new service deployment  request.

\section{Competitive analysis}
\label{sec:competitive}
In this section, we analyze FlexShare's performance in terms of the number of activated VMs, using a simplified scenario where
\begin{inparaenum}[(i)]
\item all VMs have the same maximum capacity~$C$, and
\item \label{assumption2} it is always cheaper to increase the computation capability of an existing VM than to activate a new one, i.e., $\kappa_f>C\kappa_p$.
\end{inparaenum}

As FlexShare receives a sequence of service deployment requests, each requiring multiple VNFs, we analyze the competitive ratio of the algorithm at an arbitrary point within this sequence, once FlexShare has successfully processed the VNFs of all previous service requests. 
The following lemma shows that no two instances of the same VNF, foreseen within a FlexShare solution, could be merged in a single VM. 
\begin{lemma}
Consider the deployment of the services determined through FlexShare upon receiving a new service to deploy.
Let $m_1$ and  $m_2$ be two VMs running VNF $v$ and servings sets of services $B_1$ and $B_2$, respectively. 
Then any other deployment that is identical to the one produced by FlexShare, except for the fact that $B_1\cup B_2$ are both served by the same VM, is infeasible.
\label{lem:FlexShareLoad}
\end{lemma}
\begin{IEEEproof}
We recall that sub-optimalities due to the finite services lifetime are removed in FlexShare through the procedure described in \Sec{drop}, which is run prior to every new request of service deployment. Since   $\kappa_f>C\kappa_p$,  if the procedure in \Sec{drop} ends up  de-activating any VMs, then it \emph{necessarily} reduces the cost. This implies that no two sets of service traffic flows, each set  running  on a different VM, can be moved and run on a single VM, since otherwise FlexShare would have performed the merge.
%If, at any point in time, there are two VMs~$m_1, m_2$ such that one could manage the combined load of both, then such a situation is detected and rectified by the procedure described in \Sec{drop}. We can thus guarantee that the lemma holds by running the procedure in \Sec{drop} at least once between every service de-instantiation and the first subsequent service instantiation.
\end{IEEEproof}

We now focus on the feasible deployments produced by FlexShare. These necessarily meet the target delay of all services, i.e., for any service $s$, the sojourn time associated with each VNF $v$ composing the service, is such that:
\begin{equation}
\label{eq:vnfdelay}
S(s,v)= \eta(s,v)D^{\max}(s) \quad \mbox{\em s.t.} \quad  \sum_v \eta(s,v)=1 \,.
\end{equation}
Note that in the above expression  equality holds since, otherwise, the service deployment cost could be further reduced.
 
Given any delay value~$d$ and VNF~$v\in\Vc$, we
define the {\em load gap} implied by $d$ as $\theta_v(d)=\sqrt{\frac{C_v}{d}}$, where~$C_v=\frac{C}{l(v)}$. 
The following lemma shows a sufficient (but not necessary) condition for a VM to provide a delay of at most $d$ for all services sharing a VNF $v$ running on the VM. 
\begin{lemma}[$\theta_v$-load gap]
\label{lem:epsd}
Let $v$ be a VNF that runs on VM $m$,
and let $L(m,v)$ denote the load on $m$. If the normalized computation capability ($\mu(m,v)= \mu(m)/l(v)$) satisfies
$L(m,v)+\theta_v(d)\le \mu(m,v)\le C_v$, then the sojourn time associated with any VNF $v$ composing $s$ and running on $m$, is at most $d$. 
\end{lemma}
\begin{IEEEproof}
By simple algebraic manipulation of \Eq{S}, we obtain
\begin{equation}
\label{eq:S_equiv}
\begin{array}{rcl}
S(s,v)
&=& \frac{l(v)}{\mu(m)}
\frac{1}{1-l(v)\frac{\Lambda(s,v)}{\mu(m)}}
\frac{1}{1-l(v)\frac{\Lambda(s,v)+\lambda(s,v)}{\mu(m)}} \\
&=& \frac{1}{\mu(m,v)} \frac{1}{1-\frac{\Lambda(s,v)}{\mu(m,v)}}
\frac{1}{1-\frac{\Lambda(s,v) + \lambda(s,v)}{\mu(m,v)}} \\
&=& \mu(m,v)
\frac{1}{\left(\mu(m,v) - \Lambda(s,v)\right)}
\frac{1}{\left(\mu(m,v) - \Lambda(s,v) - \lambda(s,v)\right)}.
\end{array}
\end{equation}
By definition of $L(m,v)$ and assumption on $\mu(m,v)$,
\begin{eqnarray}
\label{eq:L-dis}
\mu(m,v) - \Lambda(s,v) 
&\geq& \mu(m,v) - \Lambda(s,v) - \lambda(s,v) \\
&\geq& \mu(m,v)-L(m,v) \\
&\geq& \theta_v(d),
\end{eqnarray}
and also $\mu(m,v) \leq C_v$.

Using \ref{eq:L-dis}) into the equivalent of \Eq{S} shown above,  we obtain that the sojourn time of flows of service $s$ for VNF $v$ running on $m$ is at most
\begin{equation}
\mu(m,v)\frac{1}{\left[\mu(m,v)-\Lambda(m,v)\right]^2} 
\leq C_v \frac{1}{\theta_v(d)^2} = d,
\label{eq:eps_service_gap_implies_delay}
\end{equation}
as required.
\end{IEEEproof}

For each VNF $v$, let $d_v = \min_{s}S(s,v)$, and let $\theta_v = \theta_v(d_v)$. Intuitively, $d_v$ is the smallest sojourn time at  VNF $v$ for any of the services using $v$, and $\theta_v$ is the load gap implied by $d_v$. 
Our competitive ratio analysis consists in showing a bound to the average load on each VM activated by FlexShare (\Lem{algload}).
To this end, we define $N_v$ (resp.~$N^\star_v)$ to be the number of VMs running $v$ based on the decisions made by FlexShare (resp. the optimal decisions). Also, let $L_v$ (resp.~$L^\star_v$) denote the average load in the solution produced by FlexShare (resp. the optimal solution).

\begin{lemma}
\label{lem:algload}
If FlexShare uses more than one VM for running VNF $v$, then the average load $L_v$ over the VMs running VNF~$v$ in FlexShare is at least $\frac{C_v-\theta_v}{2}$.
\end{lemma}
\begin{IEEEproof}
% In these problems, each request $v\in s$ is an item $a_{s,v}$ whose size is $\lambda(s,v)$. Bins for VNF $v$ all have the same capacity of $C_v-\theta_v$. 
% We observe that RLA mimics the \emph{Decreasing Best-Fit}\cite{hochbaum97approximation} (DBF) bin packing algorithm. 
Consider the case where FlexShare uses more than one VM for running VNF $v$.
Assume by contradiction that the average load $L_v$ is strictly less than $\frac{C_v-\theta_v}{2}$.
Consider the two least-loaded VMs in the solution produced by FlexShare. By  assumption, one of them must have a load strictly less than $\frac{C_v-\theta_v}{2}$. 
If the sum of loads on these two VMs is less than $C_v-\theta_v$, then the services running on these two machines could have been rearranged to run on a single machine while meeting the delay constraints of all services (by the definition of $\theta_v$ and \Lem{epsd}). This would however  contradict  \Lem{FlexShareLoad}, thereby proving the claim.
\end{IEEEproof}

\Lem{algload} shows a lower bound to the load of activated VMs in FlexShare. \Thm{compratio} leverages this lower bound, and derives an upper bound on the competitive ratio of FlexShare in terms of the number of activated VMs. Since, in the simplified scenario we consider for our analysis, capacities and costs are the same for all VMs and $\kappa_f>C\kappa_p$, minimizing the number of active VMs also minimizes the cost function \Eq{obj}.
There is no guarantee that this is the case in general scenarios; however, the pricing structure~\cite{greengrass} and energy consumption~\cite{cstates} of real-world virtualized computing facilities do suggest that fixed cost indeed represents the main contribution to the total cost.

\begin{theorem}
\label{thm:compratio}
The competitive ratio of FlexShare in terms of the number of VMs activated for running VNF $v$ is~$2 + \frac{2\theta_v }{C_v-\theta_v}$.
\end{theorem}
\begin{IEEEproof}
\Lem{algload} shows that $N_v \frac{C_v-\theta_v}{2}\leq N_v L_v$ whenever at least two VMs are activated by FlexShare\footnote{Note that if FlexShare uses less than two VMs for running VNF $v$, then this is the {\em minimal} number of VMs possible for running $v$, which implies that FlexShare is optimal with respect to $v$.}.

The optimal algorithm must serve the same load as FlexShare, hence: $N_v L_v=N^\star_v L^\star_v$. Furthermore, since the load of a VM cannot exceed its maximum capability, then $N^\star_v L^\star_v\leq N^\star_v C_v$. Combining all inequalities, we have:
\begin{align}
N_v &\leq \frac{2}{C_v-\theta_v} N^\star_v L^\star_v \\
&\leq 
\frac{2 C_v}{C_v - \theta_v}N^\star_v =\frac{2(C_v+\theta_v-\theta_v)}{C_v-\theta_v}N^\star_v\\
&= \left(2 + \frac{2\theta_v }{C_v-\theta_v}\right)N^\star_v \,.
\label{eq:derivation}
\end{align}
From \Eq{derivation}, we obtain FlexShare's competitive ratio in terms of VMs activated for running VNF $v$,  i.e.,~$\frac{N_v}{N_v^\star} \leq 2 + \frac{2\theta_v }{C_v-\theta_v}$.
\end{IEEEproof}

It is interesting to observe that the ratio guaranteed by \Thm{compratio} tends to the constant value~$2$ as the maximum capability of VMs~$C$ grows -- a trend that is already in place, and is likely to continue as network equipment increases in computational capability.

\section{Numerical results}
\label{sec:results}

In this section, we describe the reference scenarios and benchmark solutions we consider (\Sec{scenario}), followed by our numerical results obtained under the synthetic and realistic scenarios (\Sec{res-synth} and \Sec{res-real}, respectively),
and by a discussion of the running times (\Sec{runtime}).

\subsection{Reference scenarios and benchmarks}
\label{sec:scenario}

\begin{table}
\caption{
Services in the synthetic scenario
    \label{tab:synth-services}
} %caption
\begin{center}
\begin{tabular}{|c|c|c|}
\hline
Service & Arrival rate [flows/ms] & Max. delay [ms] \\
\hline
$s_1$ & 2 & 10 \\
\hline
$s_2$ & 1.5 & 7.5 \\
\hline
$s_3$ & 1 & 5 \\
\hline
\end{tabular}
\end{center}
\end{table}

\begin{figure}
\centering
\includegraphics[width=.6\columnwidth]{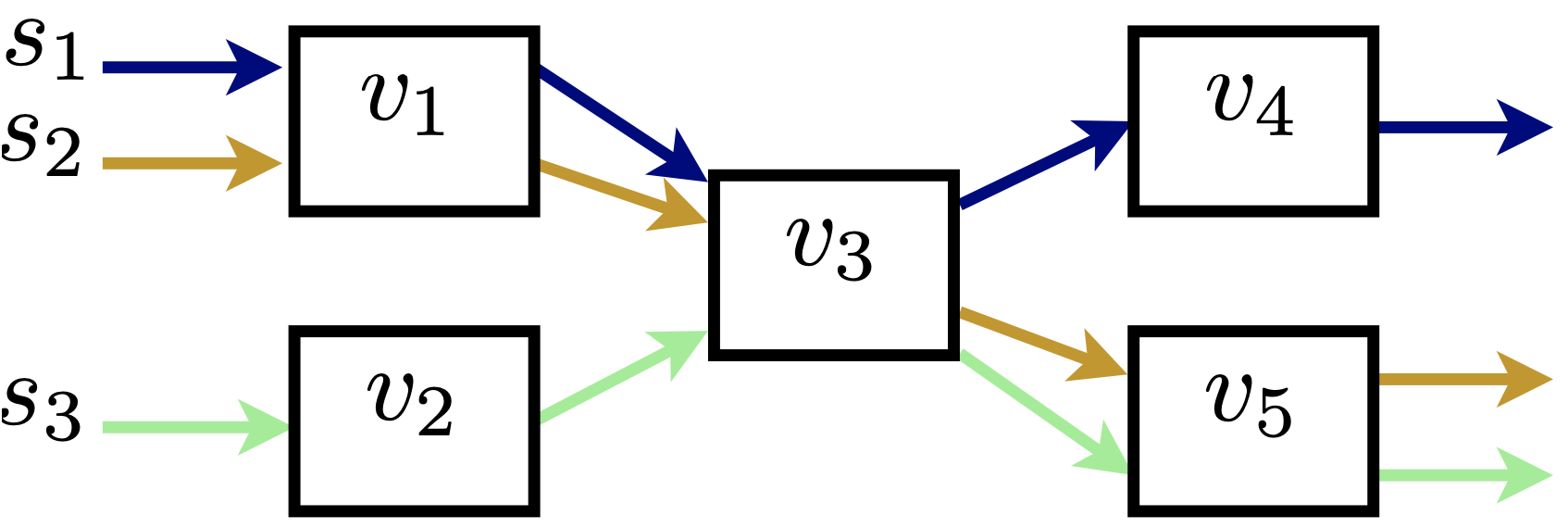}
\caption{
    VNF graphs in the synthetic scenario.
    \label{fig:synth}
} %caption
\end{figure}

\begin{figure}
\centering
\includegraphics[width=.7\columnwidth]{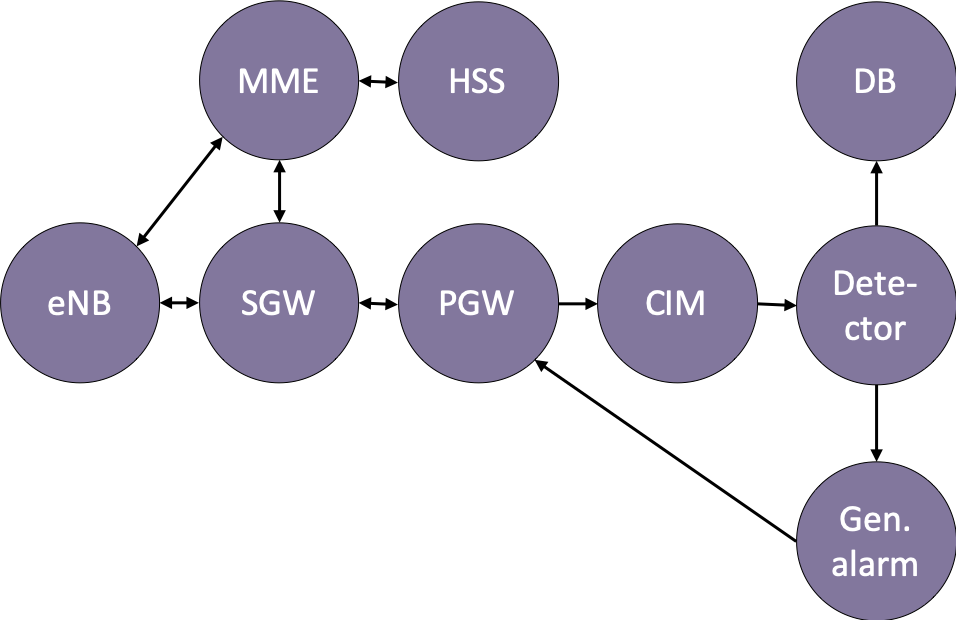}
\caption{
VNF graph of the ICA service, as described in~\cite{ica_CAMBIAPERCAMERAREADY}.
    \label{fig:vnffg-ica}
} %caption
\end{figure}

In this section, we present the two reference scenarios we consider for our performance evaluation, as well as the benchmark strategies we compare against. Without loss of generality, we consider that each service is requested exactly once, i.e., there is one service request per service.

{\bf Synthetic scenario.}
The synthetic scenario we use for performance evaluation includes three services~$s_1,\dots ,s_3$, sharing five VNFs~$v_1,\dots ,v_5$ as depicted in \Fig{synth}. All VNFs have coefficient~$l(v)=1$, while the arrival rate and maximum delay associated with each service are summarized in \Tab{synth-services}. The scenario includes~$\Mc=10$ VMs whose fixed and proportional costs are~$\kappa_f=8$ and~$\kappa_p=0.5$ units, respectively, and whose capability is randomly distributed between~5 and 10~units.

Such a scenario is small enough to allow a comparison against optimal priority assignments found by brute-force; at the same time, it contains many interesting features, including different combinations of services sharing different VNFs and different cost/capability trade-offs.

\begin{table}
\caption{
Realistic scenario: traffic flow arrival rate and computational load associated with every VNF
    \label{tab:lambdas}
} %caption
\scriptsize{
\begin{tabularx}{\columnwidth}{|X|r|r|}
\hline
{\bf VNF} & {\bf Rate~$\lambda(s,v)$} & {\bf Requirement~$l(v)$}\\
\hline\hline
\multicolumn{3}{|>{\hsize=\columnwidth}c|}{{\em Intersection Collision Avoidance (ICA)}}\\
\hline
eNB & 117.69 & $10^{-4}$\\
\hline
EPC PGW & 117.69 & $10^{-4}$\\
\hline
EPC SGW & 117.69 & $10^{-4}$\\
\hline
EPC HSS & 11.77 & $10^{-4}$\\
\hline
EPC MME & 11.77 & $10^{-3}$\\
\hline
Car information management (CIM) & 117.69 & $10^{-3}$\\
\hline
Collision detector & 117.69 & $10^{-3}$\\
\hline
Car manufacturer database & 117.69 & $10^{-4}$\\
\hline
Alarm generator & 11.77 & $10^{-4}$\\
\hline
\multicolumn{3}{|>{\hsize=\columnwidth}c|}{{\em See through (CT)}}\\
\hline
eNB & 179.82 & $10^{-4}$\\
\hline
EPC PGW & 179.82 & $10^{-4}$\\
\hline
EPC SGW & 179.82 & $10^{-4}$\\
\hline
EPC HSS & 17.98 & $10^{-4}$\\
\hline
EPC MME & 17.98 & $10^{-3}$\\
\hline
Car information management (CIM) & 179.82 & $10^{-3}$\\
\hline
CT server & 179.82 & 5 $10^{-3}$\\
\hline
CT database & 17.98 & $10^{-4}$\\
\hline
\multicolumn{3}{|>{\hsize=\columnwidth}c|}{{\em Sensing (IoT)}}\\
\hline
eNB & 50 & $10^{-4}$\\
\hline
EPC PGW & 50 &$10^{-4}$\\
\hline
EPC SGW & 50 & $10^{-4}$\\
\hline
EPC HSS & 5 & $10^{-4}$\\
\hline
EPC MME & 5 & $10^{-3}$\\
\hline
IoT authentication & 20 & $10^{-4}$\\
\hline
IoT application server & 20 & $10^{-3}$\\
\hline
\multicolumn{3}{|>{\hsize=\columnwidth}c|}{{\em Smart factory (SF)}}\\
\hline
eNB & 50 & $10^{-4}$\\
\hline
EPC PGW & 50 &$10^{-4}$\\
\hline
EPC SGW & 50 & $10^{-4}$\\
\hline
EPC HSS & 5 & $10^{-4}$\\
\hline
EPC MME & 5 & $10^{-3}$\\
\hline
Robotics control & 50 & $10^{-3}$\\
\hline
Video feed from robots & 5 & $10^{-4}$\\
\hline
\multicolumn{3}{|>{\hsize=\columnwidth}c|}{{\em Entertainment (EN)}}\\
\hline
eNB & 179.82 & $10^{-4}$\\
\hline
EPC PGW & 179.82 & $10^{-4}$\\
\hline
EPC SGW & 179.82 & $10^{-4}$\\
\hline
EPC HSS & 17.98 & $10^{-4}$\\
\hline
EPC MME & 17.98 & $10^{-3}$\\
\hline
Video origin server & 17.9 & $10^{-3}$\\
\hline
Video CDN & 179.82 & $10^{-4}$\\
\hline
\end{tabularx}
} %scriptsize
\end{table}

{\bf Realistic scenario.}
We consider five services, connected to the smart-city and smart-factory domains:
\begin{itemize}
    \item Intersection Collision Avoidance (ICA): vehicles periodically broadcast a message (e.g., CAM) including their position, speed, and acceleration; a collision detector checks if any pair of them are on a collision course and, if so, it issues an alert;
    \item Vehicular see-through (CT): cars display on their on-board screen the video captured by the preceding vehicle, e.g., a large truck obstructing the view;
    \item Urban sensing, based on the Internet-of-Things (IoT);
    \item Smart robots: a set of robots working in a factory are controlled in real-time through the 5G network;
    \item Entertainment: users consume streaming contents, provided with the assistance of a content delivery network (CDN) server.
\end{itemize}
\Tab{lambdas}, based on~\cite{pimrc-wp3,mtc,smartfactory5g}, reports the VNFs used by each service and the associated arrival rates. All services share the EPC child service, which is itself composed of five VNFs. Furthermore, the car information management (CIM) database can be shared between the ICA and the CT services.
It is worth stressing that, as exemplified in \Fig{vnffg-ica} describing the ICA service, the VNF graphs in the realistic scenarios are not simple chains but rather generic graphs.

We leverage the mobility trace~\cite{lust}, combining the real-world topology of Luxembourg City with highly realistic mobility patterns. We focus on an intersection in the downtown area, and assume that all services are deployed at an edge site located at the intersection itself. Specifically,
\begin{itemize}
    \item all vehicles within~\SI{50}{\meter} from the intersection are users of the ICA service, and send a packet (i.e., a CAM message) every~\SI{0.1}{\second};
    \item all vehicles within~\SI{100}{\meter} from the intersection are users of the CT service, and send a packet (i.e., refresh their video) every~\SI{200}{\milli\second}, i.e., the see through video has~\SI{5}{fps};
    \item those same vehicles use the entertainment video service, each consuming a~\SI{25}{fps}-video;
    \item a total of 200~sensors are deployed in the area, each generating, according to the traffic model described in the 3GPP standard~\cite{3gppmtc}, one packet every~\SI{100}{\milli\second};
    \item the smart factory contains a total of 50~robots, each requiring real-time control, and 10\% of which provides a video feed.
\end{itemize}
To tackle the most challenging scenario, we consider peak-time conditions, obtaining the request rates summarized in \Tab{lambdas}, which also reports the load~$l(v)$ associated with each VNF. As discussed in \Sec{model}, the $\lambda(s,v)$ values also incorporate the fact that not all flows visit all VNFs of a service, e.g., all ICA flows visit the local ICA server but only one in ten visits the remote one.

Finally, we assume that the PoP contains 10~VMs, each of which can be scaled up to at most~$C(m)=\SI{1000}{units}$, and each associated with fixed and proportional costs of~$\kappa_f=\SI{1000}{units}$ and $\kappa_p=\SI{1}{unit}$, respectively.

\begin{figure*}
\centering
\includegraphics[width=.3\textwidth]{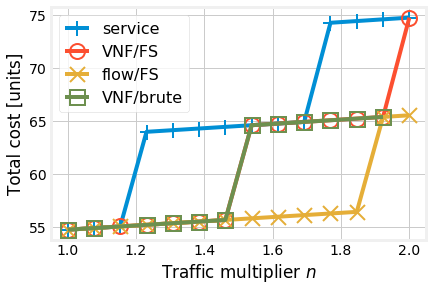}
\includegraphics[width=.3\textwidth]{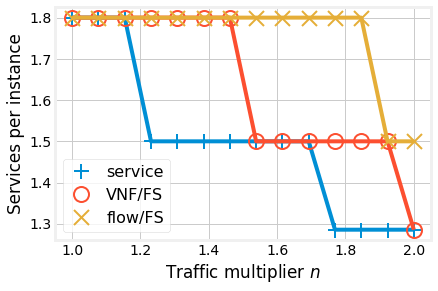}
\includegraphics[width=.3\textwidth]{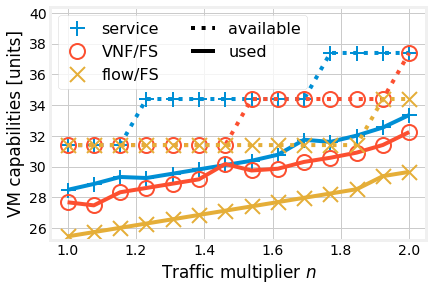}
\caption{
Synthetic scenario: total cost (left); average number of services sharing a VNF instance (center); used and maximum VM capability (right). Per-VNF and per-flow priorities are assigned via FlexShare; per-service priorities are assigned by giving higher priorities to lower-delay services.
    \label{fig:synth-details}
} %caption
\end{figure*}

{\bf Benchmark strategies.}
We study the performance of the following strategies, in increasing order of flexibility:
\begin{itemize}
    \item service-level priorities (indicated as ``{\bf service}'' in plots): priorities are assigned on a per-service basis, with lowest-delay services having the highest priority;
    \item VNF-level priorities with FlexShare (``{\bf VNF/FS}''): priorities are assigned on a per-VNF basis , and FlexShare is used to determine the VNF-level priorities~$p(s,v)$ (see Appendix~\ref{app:computelambda} and  \Sec{sub-assignment});
    \item VNF-level priorities with brute-force (``{\bf VNF/brute}''): priorities are assigned on a per-VNF basis, and all possible combinations of priorities are tested;
    \item flow-level priorities (``{\bf req./FS}''): priorities are assigned on a per-flow level, and FlexShare is used to determine them.
    %the average priorities~$\bar{p}(s,v)$.
\end{itemize}

Both FlexShare and the benchmark strategies are implemented in Python, and all tests are run on a Xeon E5-2640 server with 16~GByte of RAM.

\subsection{Results: synthetic scenario}
\label{sec:res-synth}

We start by considering the synthetic scenario and, in order to study different traffic conditions, multiply the arrival rates by a factor of~$n$, ranging between~1 and~2.

\Fig{synth-details}(left) focuses on the main metric we consider, namely, the total cost incurred by the MNO. We can observe that, as one might expect,  higher traffic translates into higher cost. More importantly, more flexibility in priority assignment results in substantial cost savings. As for per-VNF priorities, they exhibit an intermediate behavior between per-service and per-flow ones, with virtually no difference between the case where FlexShare is used to determine the priorities (``VNF/FS'') and that where all possible options are tried out in a brute-force fashion (``VNF/brute''). This highlights the effectiveness of the FlexShare strategy, which can make optimal decisions in almost all cases with low complexity.

\Fig{synth-details}(center) shows the average number of services sharing a VNF instance. It is clear that a higher flexibility in priority assignment results in more sharing, hence fewer VNF instances deployed. As $n$~increases, the number of services per instance decreases: scaling up (i.e., increasing the capability of VMs) is insufficient, and scaling out (i.e., increasing the number of VNF instances) becomes necessary.

This is confirmed by \Fig{synth-details}(right), depicting the total used VM capability (i.e.,~$\sum_{m\in\Mc}\mu(m)$) as well as the sum of the maximum values to which the capability of active VMs can be scaled up (i.e.,~$\sum_{m\in\Mc}C(m)y(m)$), denoted by solid and dotted lines, respectively. Both quantities grow with~$n$ and decrease as flexibility becomes higher. This makes intuitive sense for the maximum capability: \Fig{synth-details}(left) shows that combining FlexShare with higher-flexibility strategies results in fewer VNF instances, hence fewer active VMs. Importantly, {\em used} capability values, i.e.,~$\mu(m)$, also decrease with flexibility. Indeed, higher flexibility makes it easier to match the computational capability obtained by each service within each VNF, with its needs.

\begin{figure}
\centering
\includegraphics[width=.4\textwidth]{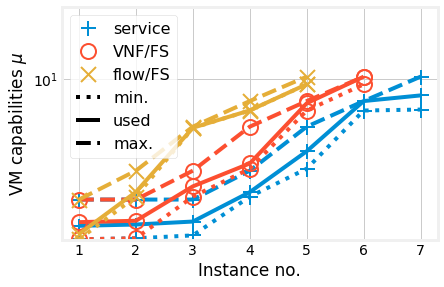}
\caption{
    Synthetic scenario, $n=1.8$: VNF capability compared to its maximum (i.e., maximum capability of the hosting VM) and minimum (i.e., required for stability) values. Per-VNF and per-flow priorities are assigned via FlexShare; per-service priorities are assigned by giving higher priorities to lower-delay services.
    \label{fig:synth-details-2}
}
\end{figure}

\Fig{synth-details-2}, obtained for~$n=1.8$, provides further insights of this phenomenon. For each VNF instance deployed by each strategy, the dotted line represents the minimum capability needed by that instance to meet target delays, the dashed one corresponds to the maximum capability~$C(m)$ that the VM can be scaled up to, and the solid line depicts the assigned capability~$\mu(m)$. We can observe that higher-flexibility strategies correspond to assigning capability values closer to the corresponding minimum, hence, less wasted capability and lower costs. Also, note how different strategies result in different numbers of created VNF instances, from~5 with flow-level priorities (the minimum possible value, as there are five VNFs) to~7 with service-level priorities. In the latter case, two instances are created for each of~$v_3$ and~$v_5$, which is not unexpected as those VNFs are shared by multiple services and hence serve higher traffic (see \Fig{synth}).

\begin{figure*}
\centering
\includegraphics[width=.3\textwidth]{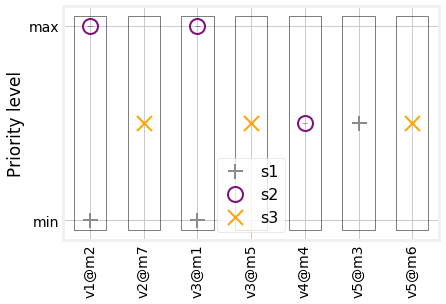}
\includegraphics[width=.3\textwidth]{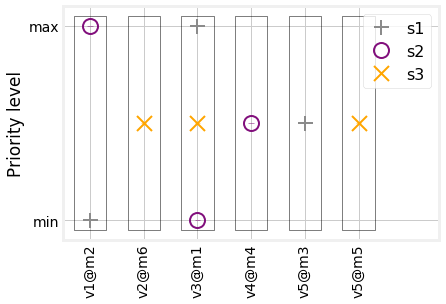}
\includegraphics[width=.3\textwidth]{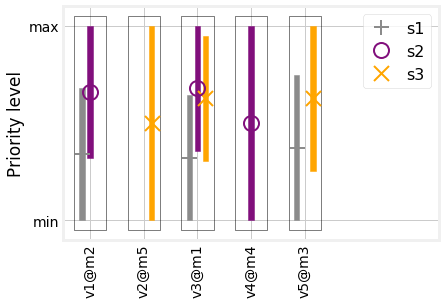}
\caption{
Synthetic scenario, $n=1.8$: priorities assigned to each service with per-service (left), per-VNF (center), and per-flow priorities (right). Per-VNF and per-flow priorities are assigned via FlexShare; per-service priorities are assigned by giving higher priorities to lower-delay services.
    \label{fig:synth-prios}
} %caption
\end{figure*}

\Fig{synth-prios} provides a qualitative view of how priorities are assigned to different services across different VNF instances. When priorities are assigned on a per-service basis (\Fig{synth-prios}(left)), services with lower target delay invariably have higher priority. If priorities are assigned on a per-VNF basis, as in \Fig{synth-prios}(center), the priorities of different services can change across VNF instances, e.g., $s_2$~has priority over~$s_1$ in the $v_1$ instance deployed at VM~$m_2$, but the opposite happens in the $v_3$ instance deployed at VM~$m_1$. \Fig{synth-prios}(right) shows that if per-flow priorities are possible, services can be combined in any way at each VNF instance.

\subsection{Results: realistic scenario}
\label{sec:res-real}

\begin{figure*}
\centering
\includegraphics[width=.3\textwidth]{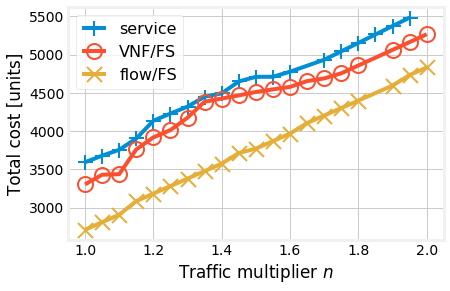}
\includegraphics[width=.3\textwidth]{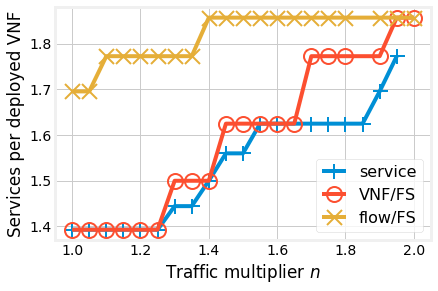}
\includegraphics[width=.3\textwidth]{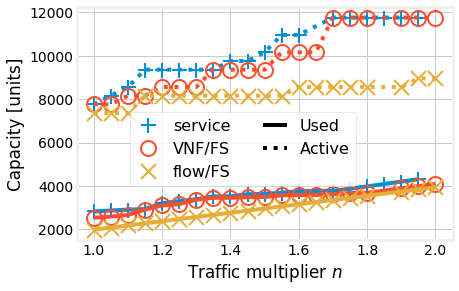}
\caption{
Realistic scenario: total cost (left); average number of services sharing a VNF instance (center); used and maximum VM capability (right). Per-VNF and per-flow priorities are assigned via FlexShare; per-service priorities are assigned by giving higher priorities to lower-delay services.
    \label{fig:real-details}
} %caption
\end{figure*}

We now move to the realistic scenario, again multiplying the arrival rates reported in \Tab{lambdas} by a factor of~$n$, varying between~1 and~2. Recall that, owing to the larger scenario size, no comparison with the brute-force strategy is possible.

\Fig{real-details}(left) shows how the total cost yielded by the different strategies has the same behavior as in the synthetic scenario (\Fig{synth-details}(left)): the higher the flexibility, the lower the cost. Furthermore, for very high values of~$n$, all strategies yield the same cost; in those cases,
%the traffic is so high that
few or no VNF instances can be shared, regardless of how priorities are assigned.

\Fig{real-details}(center) shows that VNF instances are shared among services; by comparing it to \Fig{synth-details}(center) we can observe how the behavior of per-VNF priorities tends to be closer to per-flow priorities than to per-service ones. This suggests that, even in large and/or complex scenarios, per-VNF priorities can be a good compromise between performance and implementation complexity.

\Fig{real-details}(right) shows a much larger difference between used and maximum capabilities compared to \Fig{synth-details}(right). This is due to the fact that, as can be seen from \Tab{lambdas}, there are fewer VNFs that are common among different services, and thus fewer opportunities for sharing.

\subsection{Running time}
\label{sec:runtime}

The results presented in \Sec{res-synth} and \Sec{res-real} prove that FlexShare is {\em effective}, i.e., it is able to make high-quality (indeed, often optimal) decisions. Although its worst-case computational complexity has been proven to be polynomial in \Sec{polynomial}, it is natural to wonder how long FlexShare takes to make its decisions, in the two scenarios we investigate.

The results are summarized in \Tab{runtime}: in our implementation, FlexShare never takes more than few minutes to make a decision. It is important to stress that, although such running times  are already adequate for many real-world scenarios, they can be significantly reduced. Indeed, our Python implementation of FlexShare leverages the optimization routines included in the \texttt{scipy} library, which are themselves based on decade-old FORTRAN libraries: they are adequate for prototyping and testing, but hardly a match for commercial solvers like CPLEX and Gurobi. Furthermore, as one may expect, the realistic scenario is associated with longer solution times; intuitively, this is connected with the higher number of alternatives to explore therein.

\begin{table}
\caption{Running time (in minutes) of our FlexShare implementation, in the synthetic and realistic scenarios
    \label{tab:runtime}
} %caption
\scriptsize
\begin{tabularx}{\columnwidth}{|l|X|X|}
\hline
{\bf Traffic multiplier} & {\bf Synthetic scenario} & {\bf Realistic scenario} \\
\hline
1 & 4 & 6  \\
\hline
1.2 & 5 & 7\\
\hline
1.4 & 6 & 6 \\
\hline
1.6 & 5 & 10 \\
\hline
1.8 & 7 & 9 \\
\hline
2 & 7 & 12 \\
\hline
\end{tabularx}
\end{table}

Finally, it is interesting to note that, although the running time tends to increase with the traffic, such an increase is not monotonic. This is because the running time depends on how many times the cycle represented in \Fig{flow} is executed, i.e., how many times a potentially viable deployment is found to be infeasible. This, in turn, is connected to how close to their maximum capacity VMs are, rather than to how many of them are needed.

\section{Related work}
\label{sec:relwork}

5G networks based on network slicing have attracted substantial attention, with several works focusing on 5G architecture~\cite{slicing1,slicing2}, associated decision-making issues~\cite{slicingp2,slicingalgos}, and security~\cite{orch-sec,orch-arch}.

As one of the most important decisions to make in 5G environments, VNF placement has been the focus of several studies. One popular approach is optimizing a network-centric metric, e.g., load balancing~\cite{AHirwe16} or network utilization~\cite{TKuo16}. Other papers use cost functions, e.g.,~\cite{MMechtri16,LGu16}, possibly including energy-efficiency considerations~\cite{AMarotta16,NKhoury16}. 
Recent works, e.g.,~\cite{pham2017traffic} identify energy consumption as one of the main source of operational costs (OPEX) for the MNO, and tackle it by reducing the number of idle (i.e., unused) servers.

The aforementioned works typically result in mixed-integer linear programming (MILP) models. Others cast VNF placement into a generalized assignment~\cite{infocom15_optimal}, a resource-constrained shortest path problem~\cite{martini2015latency}, or a set cover problem~\cite{eff-algos}.

Finally, a preliminary version of this work has been published in~\cite{noi-wowmom19}. Additions with respect to that version include a formal characterization of FlexShare's competitive ratio, new results obtained through real-world VNF graphs, and a more detailed description of the system model.

{\bf Novelty.}
A first novel aspect of our work is the problem we consider, i.e., VNF-sharing within one PoP as opposed to traditional VNF placement.
From the modeling viewpoint, we depart from existing works in three main ways: (i) priorities are used as a decision variable rather than as an input; (ii) different priority-assignment schemes with different flexibility are accounted for and compared; (iii) the relationship between the amount of computational resources assigned to VNFs and their performance is modeled and studied;
(iv) VM capacity scaling is properly accounted for as a necessary, complementary aspect of VNF sharing.

\section{Conclusion}
\label{sec:conclusion}

We have studied the {\em VNF sharing} problem where decision-making entities managing a single PoP have the option of sharing VNFs among several services requiring these VNFs. We have identified priority management as one of the key aspects of the problem, and found that higher flexibility in setting priorities translates into lower costs. In view of the above, we propose FlexShare, an efficient solution strategy able to make near-optimal decisions.

We have studied the computational complexity and competitive ratio of FlexShare, finding the former to be polynomial and the latter to be 
asymptotically constant as the capacity of VMs increases.
Our performance evaluation, carried out with reference to real-world VNF graphs, has highlighted how FlexShare consistently outperforms state-of-the-art alternatives, and that higher flexibility in setting priority always yields lower costs.

\bibliographystyle{IEEEtran}
\bibliography{refs}

\appendices

\section{Computing~$\Lambda(s,v)$ for relevant priority assignments}
\label{app:computelambda}

The quantity~$\Lambda(s,v)$, defined in \Sec{problem}, represents  the arrival rate of traffic flows (of any service) arriving at VNF~$v$, whose priority is higher than a generic flow of service~$s$ arriving at the same VNF~$v$. In the following, we show how such quantities can be computed under the two priority assignments discussed in \Ex{prios}, i.e., per-VNF priorities and uniformly-distributed, per-flow priorities.

\subsection{Per-VNF priorities}
\label{sec:vlevel}
We recall that, if per-VNF priorities are supported as in \Sec{example}, then all flows of each service~$s$ for VNF~$v$ are given the same deterministic priority, which we denote by~$p(s,v)$. Thus, in the per-VNF case,  $\pi(s,v)$ is always distributed according to a Dirac delta function centered in~$p(s,v)$, i.e.,~$\delta\left(\pi(s,v)-p(s,v)\right)$. Hence, $\Lambda(s,v)$ is discontinuous and given by:
\begin{equation}
\label{eq:lambda-vlevel}
\Lambda(s,v)=\sum_{t\in\Sc}H\left (p(t,v)-p(s,v)\right )\lambda(t,v),
\end{equation}
where~$H(\cdot)$ is the Heaviside step function. Indeed, intuitively a flow of service~$s$ will be queued after all flows of services~$t$ with higher priority than~$s$ (since~$H(p(t,v)-p(s,v))=1$ if~$p(t,v)>p(s,v)$), after half of the flows of services with the same priority as~$s$ (since~$H(p(t,v)-p(s,v))=0.5$ if~$p(t,v)=p(s,v)$), and before all other flows (since~$H(p(t,v)-p(s,v))=0$ if~$p(t,v)<p(s,v)$).

\subsection{Per-flow priorities}
\label{sec:rlevel}
This case corresponds to higher flexibility and  implies that priorities could follow any distribution. Below, we focus on the simple, yet relevant, case where priorities are distributed uniformly between~$r(s,v)-j$ and~$r(s,v)+j$. In this case, let us define the quantity~$q(s,t,v)=\PP(\pi(t,v)>\pi(s,v))$, whose value can be computed   through the convolution of the pdfs of~$\pi(s,v)$ and~$\pi(t,v)$. Through algebraic manipulations~\cite{diffvar} we get:
\begin{multline}
\label{eq:def-q}
\begin{aligned}
q(s,t,v) & = \PP(\pi(t,v)>\pi(s,v))\\
& = \PP\left(\pi(t,v)-\pi(s,v)>0\right)\\
\end{aligned}\\
=\begin{cases}
1 & \text{if } r(t,v){-}r(s,v)>2j\\
\frac{1}{2}{+} \frac{r(t,v)-r(s,v)}{4j} & \text{if } -2j {\leq} r(t,v){-}r(s,v){\leq} 2j\\
0 & \text{if } r(t,v){-}r(s,v)<-2j\,.
\end{cases} 
\end{multline}
Once the $q(s,t,v)$~are known, the $\Lambda(s,v)$ values can be computed by replacing \Eq{def-q} in \Eq{def-lambda}, obtaining:
\begin{equation}
\label{eq:def-lambda-rlevel}
\Lambda(s,v)=\sum_{t\in\Sc}q(s,t,v)\lambda(t,v).
\end{equation}

We can further prove that, in this case, the choice of the variation~$j$ to use in defining the variable~$\pi(s,v)$ has no influence on the possible decisions. 
\begin{property}
\label{prop:jitter}
If per-flow, uniformly-distributed priorities are used, then the choice of the variation~$j$ has no impact over the solution space.
\end{property}
\begin{IEEEproof}
The variation~$j$ only appears in the $q(s,t,v)$~quantity used in \Eq{def-q}; thus, proving the property is equivalent to showing that, if it is possible to obtain a certain value of~$q(s,t,v)$ with a certain variation~$j_1$, then it is possible to obtain the same value with any other variation~$j_2\neq j_1$.

We provide a constructive proof of this, showing that scaling all per-VNF priorities~$p(s,v)$ by~$\frac{j_2}{j_1}$ yields the same values of~$q(s,t,v)$, hence the same decisions. Focusing on the second case of \Eq{def-q}, and re-writing~$j_2$ as~$\frac{j_2}{j_1}j_1$, we have:
%\begin{equation}
%\nonumber
%\frac{\frac{j_2}{j_1}p(t,v)-\frac{j_2}{j_1}p(s,v)}{4\frac{j_2}{j_1}j_1}=
%\frac{p(t,v)-p(s,v)}{4j_1},
%\end{equation}
%and
\begin{equation}
\nonumber
\frac{\frac{2\frac{j_2+ j_2}{j_1}r(t,v)-\frac{j_2}{j_1}r(s,v)}{j_1}j_1}{4\frac{j_2}{j_1}j_1}=\frac{2j_1+r(t,v)-r(s,v)}{4j_1}.
\end{equation}
\end{IEEEproof}

\end{document}